\documentclass[aps,prb,showpacs,preprintnumbers,amsmath,floatfix]{revtex4-1}
\usepackage{graphicx}
\usepackage{dcolumn}
\usepackage{amsfonts}
\usepackage{subfigure}
\usepackage{wrapfig}
\usepackage{cancel}
\usepackage{color}
\usepackage{bm}

\newcommand{\BaFeAs}{BaFe$_{\textnormal{2}}$As$_{\textnormal{2}}$\ }

\begin{document}

\def\k{{\bf k}}
\def\rr{{\bf r}}
\def\q{{\bf q}}
\newcommand{\blue}{\textcolor{blue}}
\newcommand{\red}{\textcolor{red}}
\newcommand{\green}{\textcolor{green}}

\newcommand{\BaNiAs}{BaNi$_{\textnormal{2}}$As$_{\textnormal{2}} $}
\newcommand{\SrFeAs}{SrFe$_{\textnormal{2}}$As$_{\textnormal{2}} $}
\newcommand{\SrBaFeAs}{(Sr,Ba)Fe$_{\textnormal{2}}$As$_{\textnormal{2}}$}
\newcommand{\SrNiP}{SrNi$_{\textnormal{2}}$P$_{\textnormal{2}} $}
\newcommand{\SrKFeAs}{Sr$_{\textnormal{1-x}}$K$_{\textnormal{x}}$Fe$_{\textnormal{2}}$As$_{\textnormal{2}} $}
\newcommand{\pipi}{($\pi$,$\pi$)\ }

\title{\bf  Sensitivity of the superconducting state and magnetic susceptibility
to key aspects of electronic structure in ferropnictides}
\author{A. F. Kemper$^{1,2,5}$, T. A. Maier$^4$, S. Graser$^5$, H.-P. Cheng$^3$,  P. J. Hirschfeld$^3$,
and D. J. Scalapino$^6$}
\affiliation{$^1$Stanford Institute for Materials and Energy Science, SLAC National Accelerator Laboratory, Menlo Park, CA 64025, USA\\
$^2$Geballe Laboratory for Advanced Materials, Stanford University, Stanford CA 94305, USA\\
$^3$Department of Physics, University of Florida,
Gainesville, FL 32611, USA\\
$^4$ Center for Nanophase Materials
Sciences and Computer Science and Mathematics Division, Oak Ridge
National Laboratory, Oak Ridge, TN 37831-6494 \\
$^5$
Center for Electronic Correlations and Magnetism, Institute of Physics,
University of Augsburg, D-86135 Augsburg, Germany
\\$^6$ Department of Physics, University of California, Santa
Barbara, CA 93106-9530 USA}
\date{\today}

\begin{abstract}
Experiments on the iron-pnictide superconductors appear to show
some materials where the ground state is fully gapped, and others
where low-energy excitations dominate, possibly indicative of
gap nodes. Within the framework of a 5-orbital spin
fluctuation theory for these systems, we discuss how changes in
the doping, the electronic structure or interaction parameters
can tune the system from a fully gapped to nodal sign-changing
gap with $s$-wave ($A_{1g}$) symmetry ($s^\pm$).
In particular we focus on the role of the hole pocket at the
$(\pi,\pi)$ point of the unfolded Brillouin zone identified
as crucial to the pairing by Kuroki {\it et al.}\cite{k_kuroki_09}, and show that
its presence leads to additional nesting of hole and electron
pockets which stabilizes the isotropic $s^\pm$ state. The
pocket's contribution to the pairing can be tuned by doping,
surface effects, and by changes in interaction parameters, which
we examine. 
Analytic expressions for orbital pairing vertices calculated within the
random phase approximation
fluctuation exchange approximation allow us to draw connections
between aspects of electronic structure, interaction parameters, and the form
of the superconducting gap.
\end{abstract}

\maketitle

\section{Introduction}

\label{sec:intro}

In any new class of superconductors, the structure of the order
parameter is an important clue to the nature of the pairing
mechanism, but the determination of this structure is seldom
immediate. In the cuprates, for example, many different types of
experiments on a variety of samples had to be analyzed and compared
before a consensus was achieved, and the experimental picture was
not clarified until the effects of disorder were understood and
clean samples were prepared. It is therefore not unexpected that
the symmetry and momentum structure of the gap in the Fe-based
superconductors are still controversial nearly two years after
their discovery.\cite{y_kamihara_08} Nevertheless, the range of
behaviour seen in different materials is striking.\cite{k_ishida_09}
Here we pose, from the point of view of a weak coupling fluctuation
exchange theory,\cite{k_kuroki_08,s_graser_08} the questions: is it
possible that the superconducting state of the ferropnictides is
intrinsically sensitive to aspects of the electronic structure which
``tune" the pairing interaction? If so, which degrees of freedom
are most important?

\begin{figure}[ht]
\vspace{5mm}
\includegraphics[width=.45\columnwidth,angle=0]{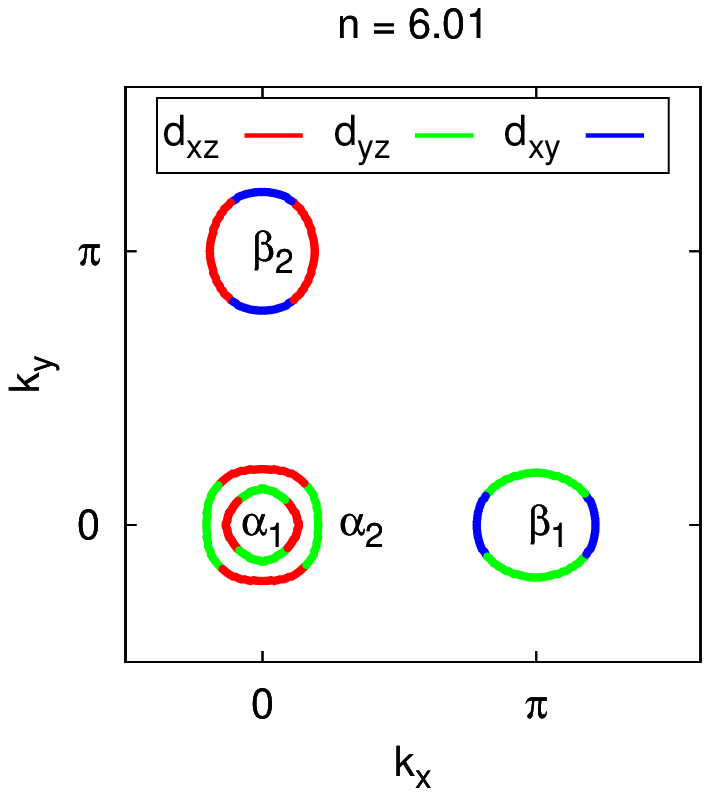}
\includegraphics[width=.45\columnwidth,angle=0]{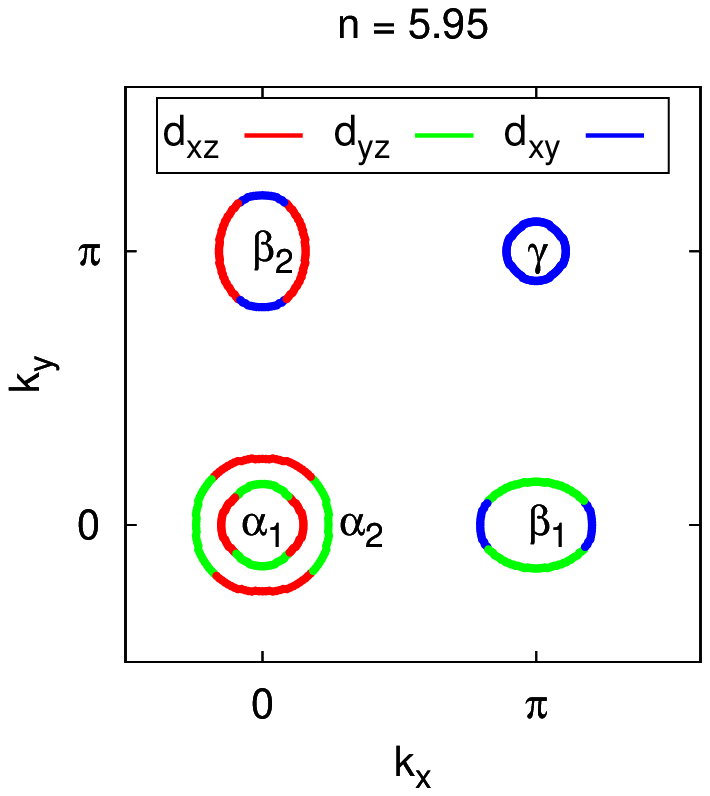}
\caption{Fermi sheets of the five-band model for $n=6.01$ (left) and  $n=5.95$ (right) with colors indicating majority orbital character
(red=$d_{xz}$, green=$d_{yz}$, blue=$d_{xy}$).  Note the $\gamma$ Fermi surface sheet is a hole
pocket which appears for $\sim 1\%$ hole doping.
}
\label{fig:fermisurface}
\end{figure}

Based on density functional theory (DFT),
\cite{s_lebegue_07,d_singh_08,c_cao_08}
quantum oscillations and angle-resolved photoemission experiments
(ARPES)\cite{l_zhao_08,h_ding_08,t_kondo_08,d_evtushinsky_09,k_nakayama_09,l_wray_08}, the Fermi surface of the Fe-pnictides is believed to
consist of a few small hole and electron pockets, as shown in
Fig.~\ref{fig:fermisurface}, where we have also indicated the
predominant Fe-orbital character of the various parts of the Fermi
surface taken from the DFT calculations of
Cao {\it et al.}~\cite{c_cao_08} for the LaFeAsO material. We will
follow the convention in Ref.~\onlinecite{s_graser_08} and elsewhere
and refer to the hole pockets around the $(0,0)$ point as the
$\alpha$ sheets and the electron pockets around the $(\pi,0)$ and
$(0,\pi)$ points of the unfolded (1-Fe) Brillouin zone as the
$\beta$ sheets.  Early on, it was proposed by
Mazin {\it et al.}~\cite{i_mazin_08}
and
Dong {\it et al.}~\cite{j_dong_08}
that the nested structure of this Fermi surface would lead to a
peak in the magnetic susceptibility near $(\pi,0)$, and that
this might drive a sign change in the superconducting order
parameter between the $\alpha$ and $\beta$ sheets~\cite{i_mazin_08}.
Several experiments on the Fe-based superconductors are indeed
consistent with a gap which is isotropic (independent
of momentum on a given pocket), but possibly with overall sign
change of this type. Angle-resolved photoemission spectroscopy
(ARPES) experiments, while not sensitive to the sign of the gap,
are the most direct measure of its magnitude, and have consistently
provided evidence taken to support an isotropic gap structure in
momentum space.~\cite{l_zhao_08,k_nakayama_09}
The observation of a resonance in inelastic
neutron scattering is strong evidence for a sign change of the
superconducting gap~\cite{a_christianson_08,m_lumsden_09,s_chi_09,t_maier_08b,m_korshunov_08a,t_maier_09a}.

On the other hand, many other experiments appear to support the
existence of low-lying excitations below the apparent gap energy.
For example, in both the LaFePO system~\cite{j_fletcher_09, c_hicks_09} and in
BaFe$_2$As$_{1-x}$P$_x$~\cite{k_hashimoto_09}, a linear-$T$ dependence
of the low-$T$ penetration depth $\Delta \lambda(T)$ has been
reported, and in Ba(Fe$_{1-x}$Co$_x)_2$As$_2$, $\Delta\lambda$
varies close to $T^2$ over most of the phase diagram~\cite{r_gordon_08};
these power laws are in contrast to the activated temperature
dependences expected for an isotropic gap. Similar power laws
have been observed in NMR\cite{r_klingeler_08,k_matano_08,h_grafe_08,k_ahilan_08,t_nakai_08,m_yashima_09}, thermal conductivity\cite{x_luo_09,m_yamashita_09,j_checkelsky_08,m_tanatar_09,y_machida_09,l_ding_09,k_hashimoto_09}, and Raman
scattering\cite{b_muschler_09}. One obvious way of interpreting these observations
is to suppose that the superconducting gap has nodes
on parts of the Fermi surface, allowing for the excitation of
quasiparticles at arbitrarily low energies. However, one may
also show that in an isotropic ``sign-changing $s$-wave"
($s^\pm$) superconductor,
disorder can create subgap
states~\cite{a_golubov_97} under certain conditions, depending on the
ratio of inter- to intraband impurity scattering. An impurity band
at the Fermi level in an $s^\pm$ state will also lead to
$\Delta \lambda \sim T^2$. There is no
known scenario for producing $\Delta \lambda \sim T$ with
impurity scattering in a gapped state, however. It is extremely
important to establish whether low-energy excitations are intrinsic
(nodal) or extrinsic (disorder-induced), and under what
circumstances fully developed gaps, as opposed to highly
anisotropic gaps, possibly with nodes, should be expected.

From the standpoint of fluctuation exchange theories of pairing
based on realistic Fermi surfaces in these materials, the most
likely states indeed appear to be preferentially of ''$s$-wave"
symmetry, with quasi-isotropic gaps on the hole pockets but highly
anisotropic states on the electron pockets~\cite{k_kuroki_08,f_wang_09,s_graser_08}.
All of these calculations indicate the proximity of other pairing
channels, particularly one with $d_{x^2-y^2}$ symmetry, but
transitions between an $s$-wave state and a $d$-wave state would
give rise to thermodynamic anomalies which have not yet been
convincingly observed. Attention has therefore focussed primarily
on the possibility of $s$-wave ($A_{1g}$ symmetry) states with
``accidental" nodes, i.e. nodes whose existence is due to details
of the pairing interaction rather than symmetry. When the leading instability was
of $s$-wave ($A_{1g}$)type, 5-orbital calculations found highly anisotropic
states for all values of parameter space explored~\cite{s_graser_08}, in apparent
contradiction to the existence of nearly isotropic states 
experimentally observed in some materials.

What aspects of the physics of these materials are responsible
for the nodes or near-nodes seen in these theories?
Some observations on this question have already been made.
Maier {\it et al.}~\cite{t_maier_09b} pointed out that, within
a model with intra- and inter-orbital interactions, nodes were
driven by the intra-orbital Coulomb repulsion, the scattering between
the two $\beta$ sheets neglected in simpler 2-band approaches,
and a tendency (observed for the parameters considered in that
work which were consistent with local spin rotational invariance)
of electrons in like orbitals to pair. Kuroki
{\it et al.}~\cite{k_kuroki_09} made an important connection between
the lattice structure, electronic structure, and pairing state of
the Fe-based superconductors, observing that in DFT calculations
the pnictide atom height above the Fe plane appeared to control
the appearance of a third $\gamma$ Fermi surface sheet centered on the
$\Gamma$ point in the folded zone corresponding to a ($\pi,\pi)$ pocket in the
unfolded zone.
This new hole-type pocket,
not considered in Ref.~\onlinecite{s_graser_08}, stabilizes a more isotropic
$s^\pm$ state. When the $\gamma$ pocket is
present, intra-orbital $q\sim(\pi,0)$ and $(0,\pi)$ scattering of $d_{xy}$
pairs between the $\gamma$ and $\beta$ pockets favor a nodeless $s^\pm$ state.

Within a model with band interactions, Vildosola {\it et al.}~\cite{v_vildosola_08} and Calderon
{\it et al.}~\cite{m_calderon_09} have also discussed the change in
the electronic structure caused by the shift of the pnictogen.
In particular, the latter authors have noted that a change in
the angle $\alpha$ formed by the Fe-As bonds and the Fe-plane
can modify the orbital content as well as the shape of the Fermi
surface sheets. In a similar model, Chubukov
{\it et al.}~\cite{a_chubukov_09} deduced a phase diagram
manifesting a transition between a nodal and fully gapped state
with $s$ symmetry as a function of a parameter controlling
the relative importance of intraband repulsion, and Thomale
{\it et al.}~\cite{r_thomale_09} reached similar conclusions
within a 4-band model, exploring the stability of the nodeless
state with respect to doping and other changes in electronic
structure.  
Ikeda {\it et al.}~\cite{h_ikeda_10} considered the doping dependence of spin fluctuations
and electron correlations within the renormalized 
fluctuation exchange (FLEX) approximation using LDA dispersions;
they also observe anisotropic behaviour
of the gap function around the M point for the electron doped material, and fully-gapped
behaviour in the hole-doped material. In a subsequent work, they explored the doping
dependence of the pairing state within FLEX.~\cite{h_ikeda_10a}
Wang {\it et al.}~\cite{f_wang_10} have also discussed the important role played by the
$\gamma$ Fermi surface and emphasized the role of the orbital matrix elements
in determining the momentum structure of the gap. From their functional
renormalization group calculations, they find that the degree of gap anisotropy
depends upon the relative weight of the orbital matrix elements. They find
that by tuning some of the tight-binding parameters to alter the relative
orbital weights, one can change the anisotropy of the gap. Thomale {\it et al.}~\cite{r_thomale_10}
have also recently reported functional renormalization group results and argue that a
nodal superconducting gap can appear on the $\beta$ electron Fermi surfaces when
the $\gamma$ Fermi surface is absent. 

Many authors have considered a Hamiltonian with all possible two-body on-site
interactions between electrons in Fe orbitals as a good starting
point for a microscopic description of this system,
\begin{eqnarray}
H & = & H_{0}+\bar{U}\sum_{i,\ell}n_{i\ell\uparrow}n_{i\ell\downarrow}+\bar{U}'\sum_{i,\ell'<\ell}n_{i\ell}n_{i\ell'}\\
 &  & +\bar{J}\sum_{i,\ell'<\ell}\sum_{\sigma,\sigma'}c_{i\ell\sigma}^{\dagger}c_{i\ell'\sigma'}^{\dagger}c_{i\ell\sigma'}c_{i\ell'\sigma} \nonumber \\
 &  & +\bar{J}'\sum_{i,\ell'\neq\ell}c_{i\ell\uparrow}^{\dagger}c_{i\ell\downarrow}^{\dagger}c_{i\ell'\downarrow}c_{i\ell'\uparrow} \nonumber
\label{H}
\end{eqnarray}
where $n_{i\ell} =  n_{i,\ell\uparrow} + n_{i\ell\downarrow}$.  The
interaction parameters ${\bar U}$, ${\bar U}'$, ${\bar J}$, and
${\bar J}'$ correspond to the notation of Kuroki
{\it et al.}~\cite{k_kuroki_08}, and are related to those used by
Graser {\it et al.}~\cite{s_graser_08} by $\bar U=U$, $\bar J=J/2$,
$\bar U'=V+J/4$, and $\bar J'=J'$. The kinetic energy $H_0$ is
described by a tight-binding model\cite{s_graser_08} spanned by the
5 Fe $d$ orbitals,
which, depending on the tight binding parameters and the filling
$n$ give rise to the Fermi surfaces shown in
Fig.~\ref{fig:fermisurface}. Here in the 1 Fe per unit cell
Brillouin zone that we will use, there are the usual $\alpha_1$
and $\alpha_2$ hole Fermi surfaces around the $\Gamma$ point, the
two $\beta_1$ and $\beta_2$ Fermi surfaces around $(\pi,0)$ and
$(0,\pi)$ and for the hole doped case $n = 5.95$ shown in Fig.~\ref{fig:fermisurface}
there is an extra hole FS $\gamma$ around the $(\pi,\pi)$ point.
We use a notation in which a given orbital index $\ell\in(1,2,\ldots,5)$
denotes the Fe-orbitals ($d_{xz},d_{yz},d_{xy},d_{x^2-y^2},d_{3z^2-r^2}$).
An important role will be played by the orbital matrix elements $a^\ell_\nu(\k)=
\langle\ell|\nu \k\rangle$ which relate the orbital and band states. The
dominant orbital weights $|a^\ell_\nu(\k)|^2$ on the Fermi surfaces are illustrated
in Fig.~\ref{fig:fermisurface}.

In Eq.~(\ref{H}), we have separated the intra- and inter-orbital
Coulomb repulsion, as well as the Hund's rule exchange $J$ and
``pair hopping" term $J'$ for generality, but note that if they
are generated from a single two-body term with spin rotational
invariance they are related by $\bar U'=\bar U - 2 \bar J$ and
$\bar J'=\bar J$. Below we also consider the case where
spin rotational invariance (SRI) is explicitly broken by other
interactions in the crystal, such that $\bar J'$ and $\bar{U}'$
can take on independent values.

The analysis of individual orbital contributions to pair scattering can be a powerful
tool to understand the influence of electronic structure on pairing.  The basic picture
which emerges from the spin fluctuation theories\cite{k_kuroki_08,s_graser_08} is straightforward. 
The intra-orbital scattering of $d_{xz}$ and
$d_{yz}$ pairs between the $\alpha$ and $\beta$ Fermi surfaces by $(\pi,0)$
and $(0,\pi)$ spin fluctuations leads naturally to a gap which changes sign
between the $d_{xz}/d_{yz}$ parts of the $\alpha$ Fermi surfaces and the $d_{xz}/d_{yz}$
parts of the $\beta_1$ and $\beta_2$ electron pockets. However,
intra-orbital scattering between the $d_{xy}$ portions of the $\beta_1$ and $\beta_2$
Fermi surfaces suppresses the gap in the $d_{xy}$ regions of the $\beta$ Fermi surface
and competes with the formation of an isotropic $s^\pm$ gap. In addition, this anisotropy
can also be driven by the need to reduce the effect of the on-site Coulomb
interaction.\cite{t_maier_09b, a_chubukov_09} There can also be inter-orbital pair scattering, such that for example a
$d_{xz}$ pair on $\alpha_1$ scatters to a $d_{xy}$ pair on $\beta_1$. These
inter-orbital scattering processes which depend upon $\bar U'$ and $\bar J'$
are weaker than the intra-orbital processes for spin rotationally invariant
(SRI) parameters where $\bar U'+\bar J'=\bar U-\bar J$ and $\bar J' = \bar J$.
However, as noted by
Zhang {\it et al.}~\cite{j_zhang_09}, the interaction parameters in the solid will be
renormalized and one may have non spin rotationally invariant (NSRI) interaction
parameters leading to enhanced inter-orbital pairings.

Here we explore these issues using results obtained from a 5-orbital RPA
calculation to further examine the effects of the electronic structure, the
doping and the interaction parameters on the structure of the gap. Analytical results 
for orbital components of the pairing vertices have allowed us to understand the 
factors which influence the structure of the gap, and in particular the transition
from nodal to nodeless behavior. In
Section~\ref{sec:2} we introduce the effective pair scattering vertex, its
fluctuation-exchange RPA form and the pairing strength functional that determines
the momentum dependence of the gap function.\cite{s_graser_08} Results for
the gap function $g(\k)$ obtained for a typical set of spin rotationally
invariant (SRI) interaction parameters are discussed in Section~\ref{sec:3}.
The strongest pairings occur in the $A_{1g}$ channel and the discussion in this
section focuses on the role of the doping and the presence of the $\gamma$
pocket in determining whether the gap is nodeless. The nodeless character is
found to be related to the additional resonant pairing provided by the $\gamma$ sheet,
and is associated with the strong tendency for pairing between like orbitals
when the $\gamma$ pocket is of $d_{xy}$ character. The origins of the tendency
to pair in like orbitals
are then explored in Sec.~\ref{sec:3}. Non spin rotationally invariant (NSRI)
interaction parameters are discussed in
Section~\ref{sec:4}, where we examine the effect of the Hund's rule exchange
 $\bar J$ and the pair hopping $\bar J'$
on the pairing interaction and the  nodal structure of the gap. Section~\ref{sec:5} contains a further
discussion of the effect of the orbital character of the $\gamma$ hole
pocket as well as the possible role of surface effects in causing nodeless
behavior. Our conclusions are contained in Section~\ref{sec:6} and in the
appendix we discuss some approximate forms for the pairing vertex.

\section{Spin fluctuation pairing}
\label{sec:2}

\begin{figure}[ht]
\vspace{5mm}
\includegraphics[width=0.9\columnwidth,angle=0]{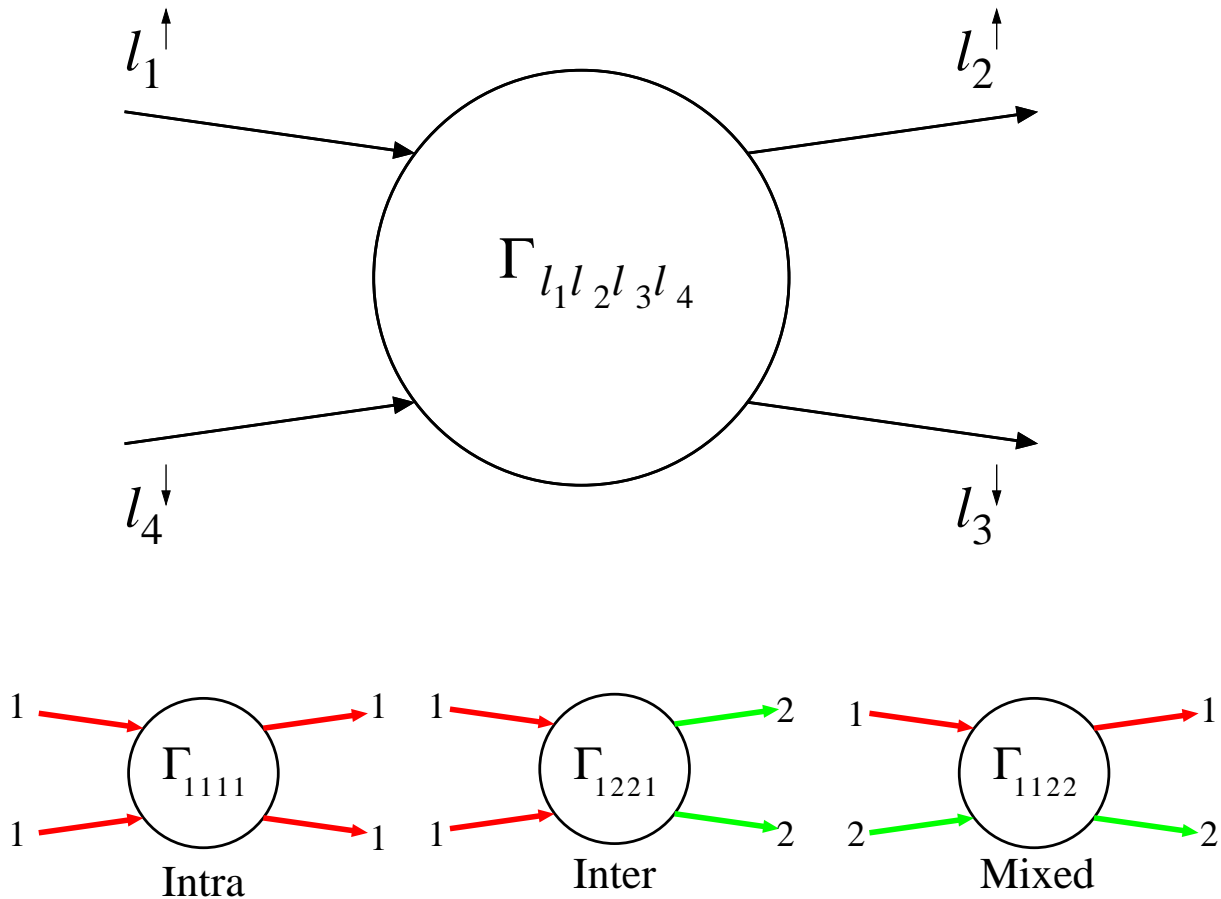}
\caption{Top: pairing vertex $\Gamma_{\ell_1\ell_2
\ell_3\ell_4}$ defined in terms of orbital states $\ell_i$ of incoming
and outgoing electrons.  Bottom: representative examples of
classes of orbital vertices referred to
in the text: intra-, inter- and mixed orbital vertices.}
\label{fig:vertex}
\end{figure}

As in Ref.~\onlinecite{s_graser_08}, we analyze the effective pair
scattering vertex $\Gamma(\k,\k')$ in the singlet channel,
\begin{eqnarray}
{\Gamma}_{ij} (\k,\k') & = & \mathrm{Re}\left[\sum_{\ell_1\ell_2\ell_3\ell_4} a_{\nu_i}^{\ell_2,*}(\k) a_{\nu_i}^{\ell_3,*}(-\k) 
 \times {\Gamma}_{\ell_1\ell_2\ell_3\ell_4} (\k,\k',\omega=0) a_{\nu_j}^{\ell_1}(\k')  a_{\nu_j}^{\ell_4}(-\k') \right] \nonumber
\label{eq:Gam_ij}
\end{eqnarray}
where the momenta $\k$ and $\k'$ are restricted to the various
Fermi surface sheets with $\k \in C_i$  and $\k' \in C_j$.  The
orbital vertex functions $\Gamma_{\ell_1\ell_2\ell_3\ell_4}$ represent the
particle-particle scattering of electrons in orbitals $\ell_1,\ell_4$ into $\ell_2,\ell_3$
(see Fig.~\ref{fig:vertex}) and in the fluctuation exchange
formulation~\cite{n_bickers_89,k_kubo_07} are given by
\begin{eqnarray}
&&{\Gamma}_{\ell_1\ell_2\ell_3\ell_4} (\k,\k',\omega) = \left[\frac{3}{2} \bar U^s
\chi_1^{\rm RPA}  (\k-\k',\omega) \bar U^s +
 \frac{1}{2} \bar  U^s
 - \frac{1}{2}\bar U^c  \chi_0^{\rm RPA}  (\k-\k',\omega)
\bar U^c + \frac{1}{2} \bar U^c \right]_{\ell_3\ell_4\ell_1\ell_2},
\label{eq:fullGamma}
\end{eqnarray}
where each of the quantities $\bar U^s$, $\bar U^c$, $\chi_1$, etc.
represent matrices in orbital space as specified in the appendix.
Note that the $\chi_1^{\rm RPA}$ term describes the spin-fluctuation
contribution and the $\chi_0^{\rm RPA}$ term the orbital
(charge)-fluctuation contribution.

For the parameter regions we will discuss, the dominant contribution
to the pairing comes from the $S=1$ particle-hole exchange given
by the first term in Eq.~(\ref{eq:fullGamma}).
The forms of the interaction
matrices $\bar U^s$ and $\bar U^c$ are given in the appendix.
As illustrated in Fig.~\ref{fig:vertex} there are intra-orbital, inter-orbital, and
mixed-orbital pair scattering processes.  The contributions of
each to the total pair scattering vertex $\Gamma_{ij}$ in Eq.~(\ref{eq:Gam_ij}) are quite different.  In particular, as discussed below,
the orbital matrix elements for $\k$ and $-\k$ states on the Fermi
surface favor pairs which are formed from electrons in the same
orbital state. We therefore find that, in spite of the fact that the mixed-orbital
scattering can be significant, its contribution to the pairing
interaction is negligible.

If one writes the superconducting gap
$\Delta(\k)$ as $\Delta g(\k)$, with $g(\k)$ a dimensionless function
describing the momentum dependence of the gaps on the Fermi
surfaces, then $g(\k)$ is determined as the stationary solution of
the dimensionless pairing strength functional~\cite{s_graser_08}
\begin{equation}
\lambda [g(\k)] = - \frac{\sum_{ij} \oint_{C_i} \frac{d k_\parallel}{v_F(\k)} \oint_{C_j}
\frac{d k_\parallel'}{v_F(\k')} g(\k) {\Gamma}_{ij} (\k,\k')
g(\k')}{ (2\pi)^2 \sum_i \oint_{C_i} \frac{d k_\parallel}{v_F(\k)} [g(\k)]^2 }
\label{eq:pairingstrength}
\end{equation}
with the largest coupling strength $\lambda$. Here the momenta $\k$
and $\k'$ are restricted to the various Fermi surfaces $\k \in C_i$
and $\k' \in C_j$ and $v_{F,\nu}(\k) = |\nabla_\k E_\nu(\k)|$ is the
Fermi velocity on a given Fermi surface. The eigenvalue $\lambda$
provides a dimensionless measure of the pairing strength.

\section{Spin rotational invariant case}
\label{sec:3}

Below we discuss the essential physics of the gapped-nodal transition
within the constrained interaction parameter subspace where spin rotational invariance
is assumed. It is worthwhile recalling the main points of the argument for
an isotropic $s^\pm$ state: 1) that a repulsive effective interaction peaked near
$(\pi,0)$ would drive such a state provided 2) this interaction did not
vary significantly over the small Fermi surface pockets.\cite{i_mazin_08}  Our main points are as follows:

\begin{itemize}

  \item The largest pair scattering processes tending to stabilize an isotropic $s^\pm$ state
        are the {\it intra}-orbital scattering pairing vertices $\Gamma_{aaaa}$ (c.f. Fig.~\ref{fig:vertex}), which are peaked near $(\pi,0)$.
  \item The intra-orbital processes primarily affect the gap on the sections of the Fermi surface with the corresponding
        orbital character. 
        The relative signs between various orbital sections are determined by subdominant intra- and  inter-orbital processes.
  \item The isotropic $s^\pm$ state can be frustrated by intraband Coulomb scattering and by pair scattering
        processes between the two electron sheets with $\q \sim (\pi,\pi)$, both of which favor nodes.\cite{t_maier_09b}
  \item The $\gamma$ pocket of $d_{xy}$ character can overcome this frustration and stabilize the nodeless state.\cite{k_kuroki_09}
  \item If the Hund's rule coupling $\bar J$ is weak,   the processes induced by the $\gamma$ pocket are not sufficient to eliminate the nodes.
        The Hund's rule exchange is necessary to overcome an  attractive $(\pi,0)$ interaction between $d_{xz}$  
        ($d_{yz}$) pairs and $d_{xy}$ pairs  and drive a strong intra-orbital $d_{xz}$ and $d_{yz}$ repulsion.

\end{itemize}

\begin{figure}[ht]
\includegraphics[width=\columnwidth]{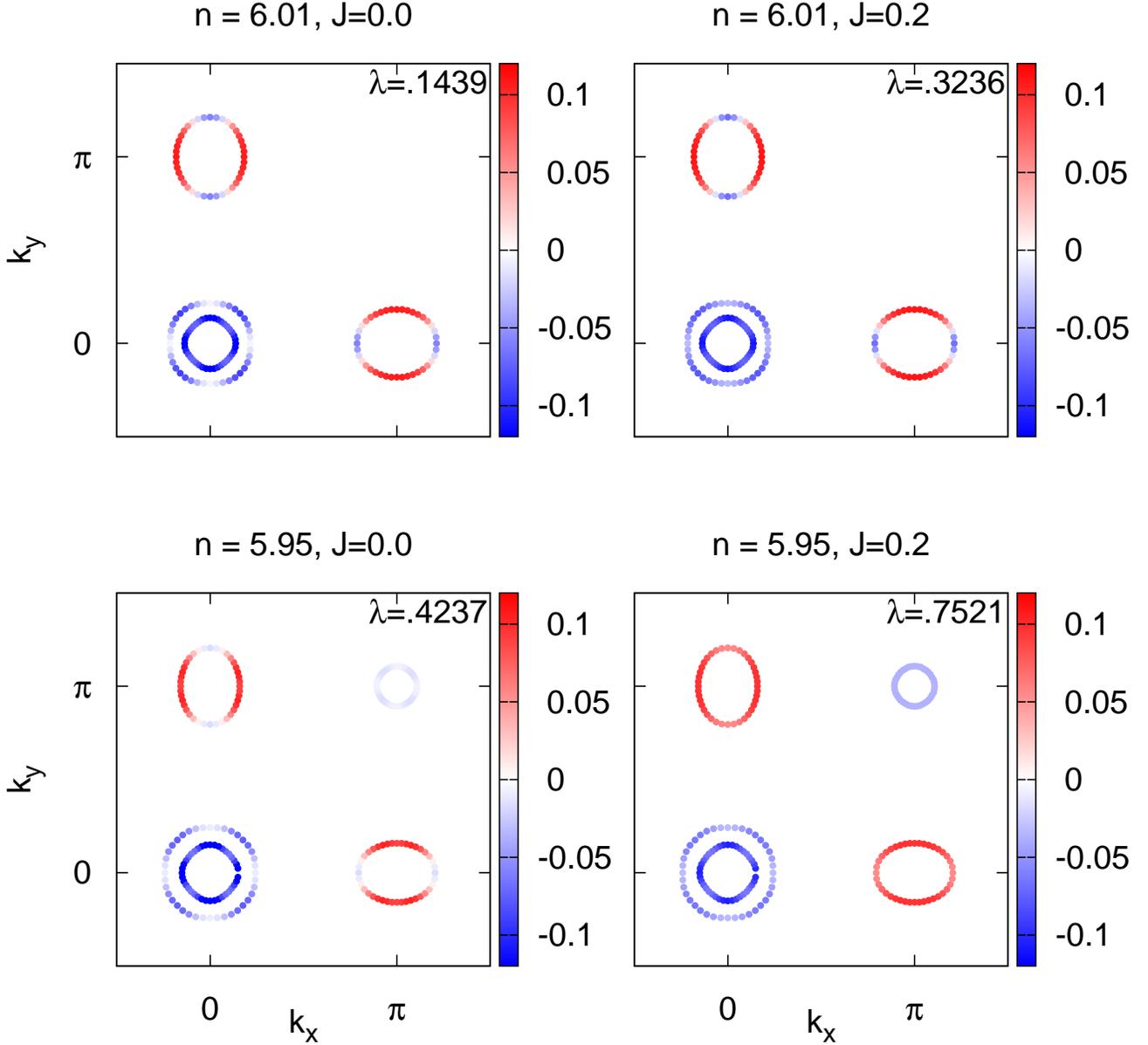}
\caption{The gap eigenfunctions $g(k)$ for a spin rotationally invariant
parameter set $\bar U=1.3$, $\bar J= 0.0, 0.2$, 
for dopings $n=6.01$ (top) and $n=5.95$ (bottom).}
\label{fig:eigenvectors}
\end{figure}

\begin{figure}[ht]
\vspace{5mm}
\includegraphics[width=0.9\columnwidth]{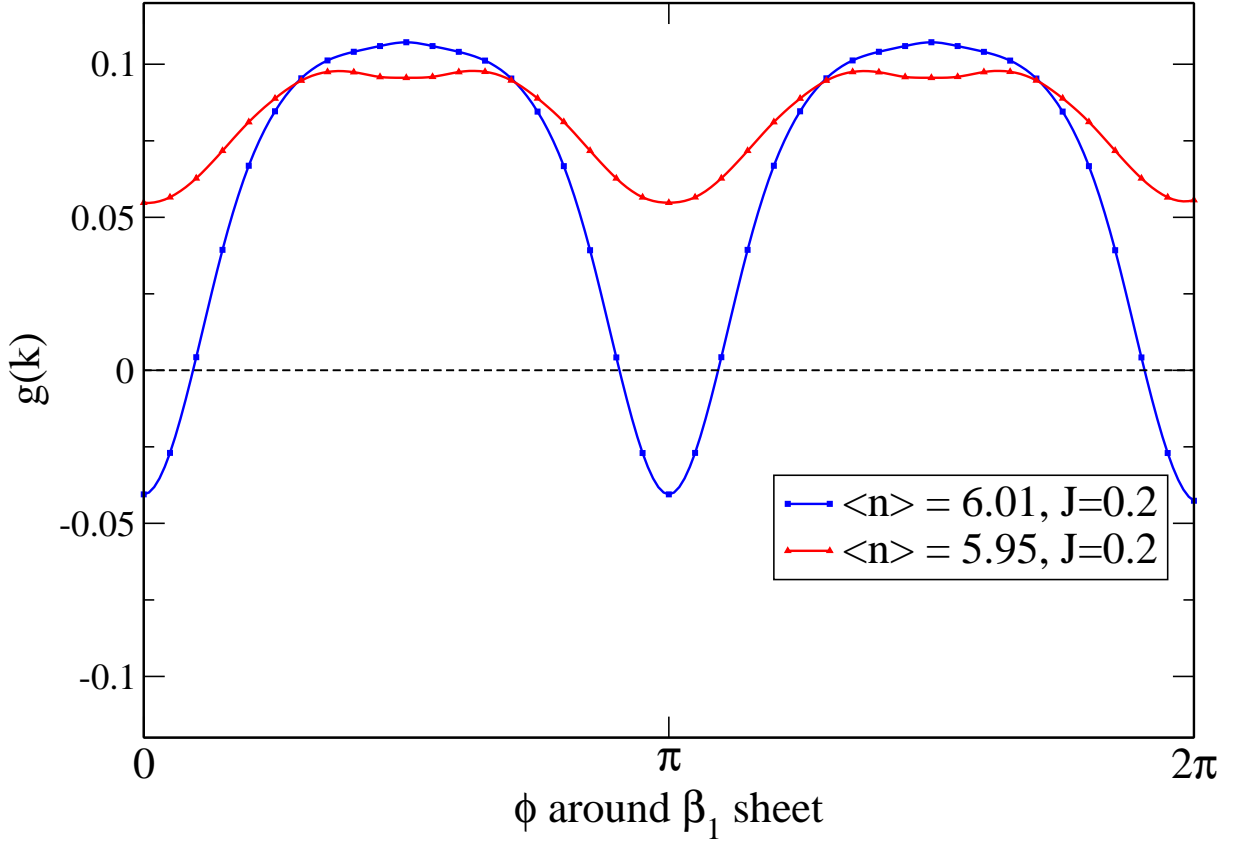}
\caption{The gap function $g(k)$ on the $\beta_1$ pocket for
$n=5.95, \bar J=0.2$ (red triangles) and
$n=6.01, \bar J=0.2$ (blue squares).
Here the angle $\phi$ is measured from the $k_x$-axis.}
\label{fig:evector_vs_angle}
\end{figure}

To illustrate these points, we begin by plotting gap functions $g(\k)$ for two typical sets of SRI interaction parameters
$\bar U=1.3$, $\bar J=0.0, 0.2$ and two different fillings
$n=6.01$ and $5.95$, as shown in Fig.~\ref{fig:eigenvectors}.
In the electron-doped case ($n = 6.01$) the $\gamma$ Fermi
surface around $(\pi,\pi)$ is absent and one finds an anisotropic
gap with nodes on the $\beta_1$ and $\beta_2$ Fermi surfaces for zero
and for finite $\bar J$.\cite{s_graser_08}
The doping dependence at fixed $\bar J$ of the gap $g(\k)$ versus angle around the $\beta_1$ Fermi surface is shown
in Fig.~\ref{fig:evector_vs_angle}.
The nodes arise in part because of the $\beta_1-\beta_2$ pair
scattering and other aspects of the interactions which frustrate the
isotropic $s^\pm$ state.\cite{t_maier_09a} In particular, the sign change of
the gaps on the end regions of the
$\beta$ Fermi surfaces reduces some of the frustration which arises
from the $\beta_1-\beta_2$ scattering.\cite{k_kuroki_09} In addition,
this sign change acts to suppress the effect of the intra-band
Coulomb repulsion.

For the hole doped case ($n=5.95$) and finite $\bar J$, the gap function,
as seen in Figs.~\ref{fig:eigenvectors} and \ref{fig:evector_vs_angle},
is anisotropic but nodeless.
As discussed in Ref.~\onlinecite{k_kuroki_09}, the scattering between
the $\gamma$ $(\pi,\pi)$ Fermi surface and the $\beta_1$ and
$\beta_2$ Fermi surfaces stabilizes the nodeless gap. While there
is still frustration associated with the $\beta_1-\beta_2$ pair
scattering, the additional $\beta_1-\gamma$ and $\beta_2-\gamma$
pair scattering processes overcome it.
The presence of the pocket also increases the overall pairing strength,
which can be seen by comparing the pairing eigenvalues in the Figure; this
would correspond to an increase in the critical temperature.
\begin{figure}[ht]
\includegraphics[width=0.50\columnwidth]{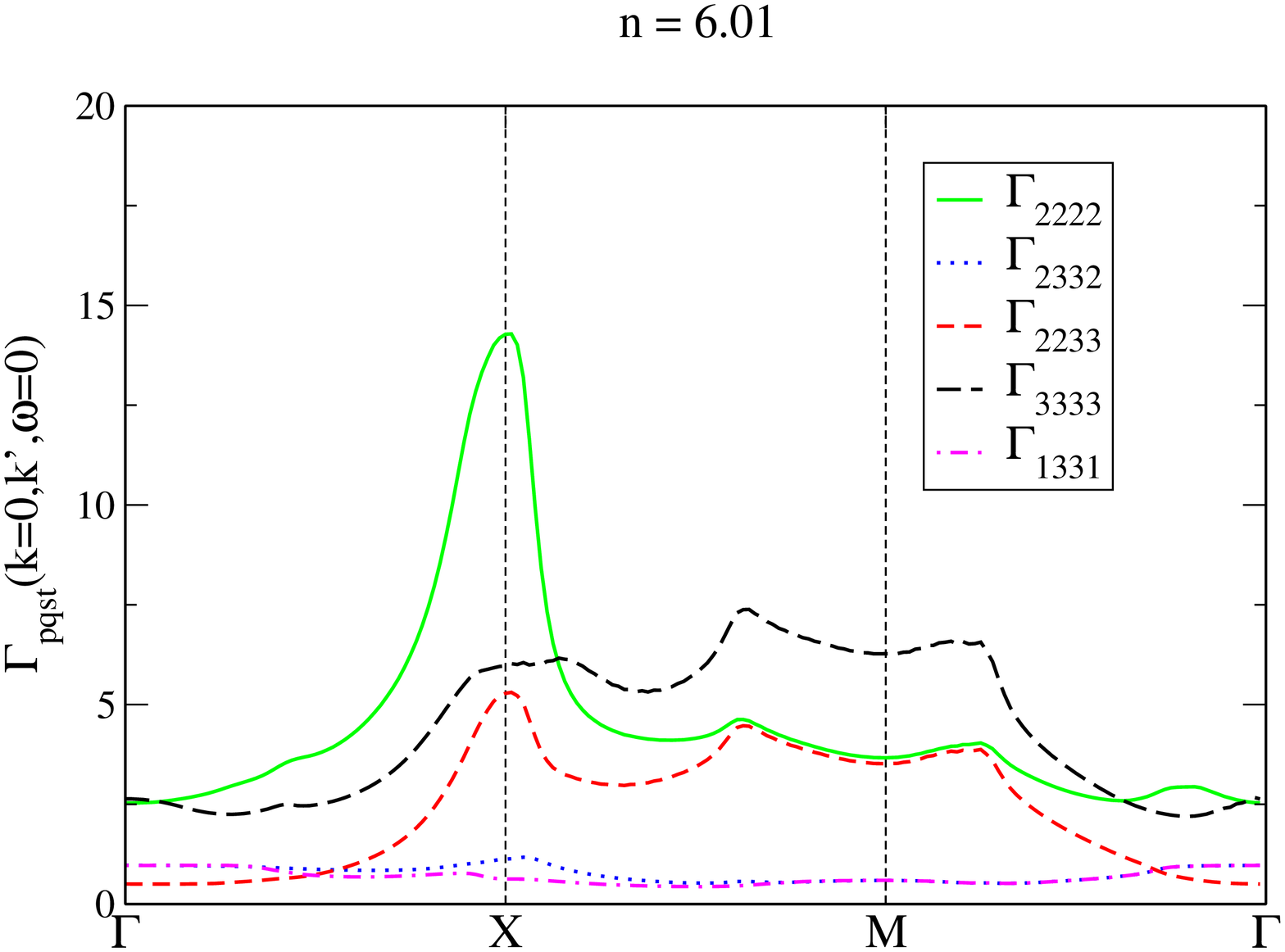}
\includegraphics[width=0.50\columnwidth]{gamma_cut_n5.95_J0.2.eps}
\caption{Orbital pairing vertices along high symmetry directions in $\bf q$
space for $\bar U=1.3$, and $\bar J=0.2$  for $n=5.95$ (bottom) and
$n=6.01$ (top),
spin rotation invariance assumed.  Solid (green) line:
$\Gamma_{2222}$ (intra); dotted (blue) $\Gamma_{2332}$ (inter);
dashed (red) $\Gamma_{2233}$ (mixed); dashed (black) $\Gamma_{3333}$;
dash-dotted (magenta) $\Gamma_{1331}$ (inter).  Note that the vertical scales in
the two panels are different.
}
\label{fig:orbital_vertices}
\end{figure}
To see how these effects arise in more detail, in Fig.~\ref{fig:orbital_vertices}
we have plotted several of the orbital pairing vertices
$\Gamma_{\ell_1\ell_2\ell_3\ell_4}(\q)$ for $\q$ along high symmetry
directions in the Brillouin zone for both $n=5.95$ and $6.01$.
The peak in  $\Gamma_{2222}(\q)$ near the $X$ point arises from
$\q=(\pi,0)$ scattering processes in which $d_{yz}$ pairs are scattered
between the $\alpha$ and $\beta_1$ Fermi surfaces. In the presence of the
$\gamma$ sheet, both $\Gamma_{3333}(\q)$ intra-orbital
$d_{xy}$ pair scattering between $\beta_{1,2}$ and $\gamma$ as well as
$\Gamma_{1331}(\q)$ and $\Gamma_{2332}(\q)$ inter-orbital $d_{xz},d_{yz}$ pair to $d_{xy}$ pair
scattering between $\beta_{1,2}$ and $\gamma$ are also possible.
The pair scattering from the $\beta$ sheets to the $\gamma$ hole pocket, represented by $\Gamma_{3333}$, provides a
strong resonant stabilization of the nodeless state; we have verified that the nodal behaviour is recovered
when it is neglected.
As discussed previously, although mixed-orbital vertices such as $\Gamma_{2233}$
are present, their contribution to $\Gamma_{ij}(\k,\k')$ is suppressed
 by the orbital matrix elements in Eq.~(\ref{eq:Gam_ij}), and the pairing
is dominated by intra- and inter-orbital scattering.

\begin{figure}[ht]
\includegraphics[width=0.45\columnwidth]{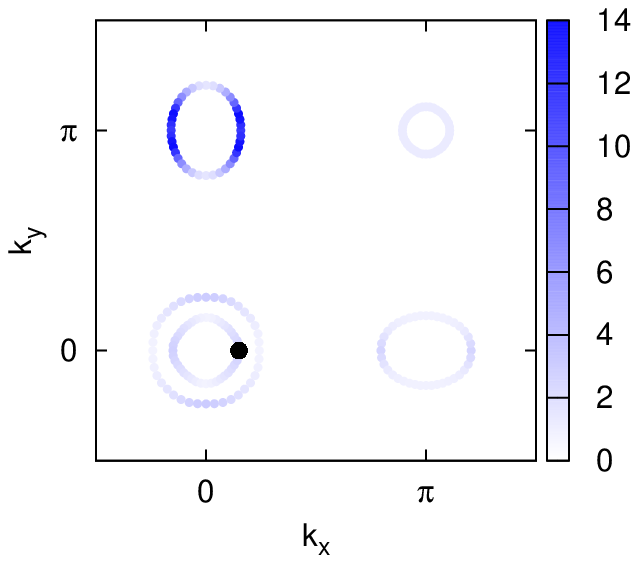}
\includegraphics[width=0.45\columnwidth]{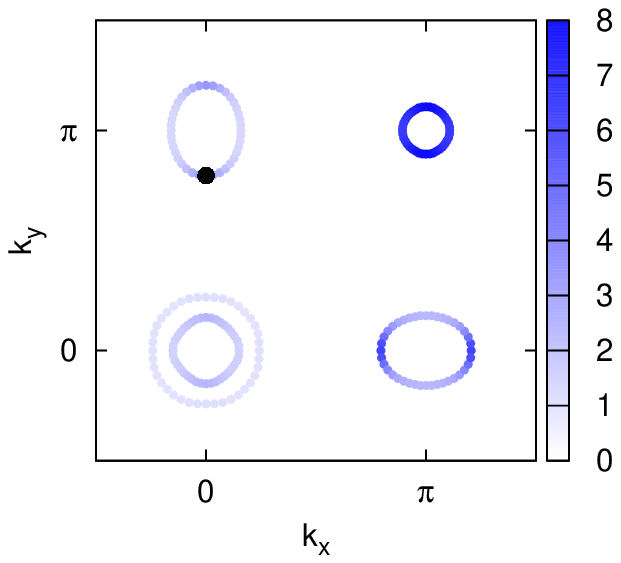}

\caption{The total pair scattering vertex $-\Gamma_{ij}(k,k')$
for $n=5.95$ with parameters $\bar U=1.3$ and $\bar J=0.2$
as a function of $k$ with $k'$ set to the point on the Fermi
surface indicated in each panel by the black dot.}
\label{fig:fullvertex}
\end{figure}

To further illustrate this, in Fig.~\ref{fig:fullvertex} we have plotted the pairing
interaction $\Gamma_{ij}(\k,\k')$ which determines the
gap via Eq.~(\ref{eq:pairingstrength}). In these plots, one member of
a $(\k',-\k')$ pair is located at the wave vector denoted by a black circle
in the figure. The plot shows the strength of the pairing
interaction $\Gamma_{ij}(\k,\k')$ associated with the scattering of
this $(\k',-\k')$ pair to a $(\k,-\k)$ pair on the various Fermi
surfaces. One sees that if the initial pair is located in a
region which has predominantly $d_{xz}$-orbital character (top), the
strongest scattering is to a pair $(\k,-\k)$ on other $d_{xz}$
regions.\cite{t_maier_09a} Similarly, there are strong $d_{yz}$ intra-orbital
scattering processes which are obtained by rotating the figures
by $\pi/2$. While inter-orbital scattering processes are also
present, they are weaker for the parameters we have considered here,
as seen, e.g. from the plot of  $\Gamma_{2332}$ in Fig.~\ref{fig:orbital_vertices}.
In addition to the $d_{yz}$ intra-orbital scattering, there is strong
$d_{xy}$ intra-orbital scattering between $\beta_2$ and $\gamma$
as well as between $\beta_1$ and $\gamma$.
Note that in Fig.~\ref{fig:fullvertex} the pairing strength is anisotropic
along the Fermi sheets, violating one of the key assumptions
underlying the argument for an isotropic $s^\pm$ state.

\begin{figure}[ht]
\vspace{15mm}
\includegraphics[width=0.9\columnwidth]{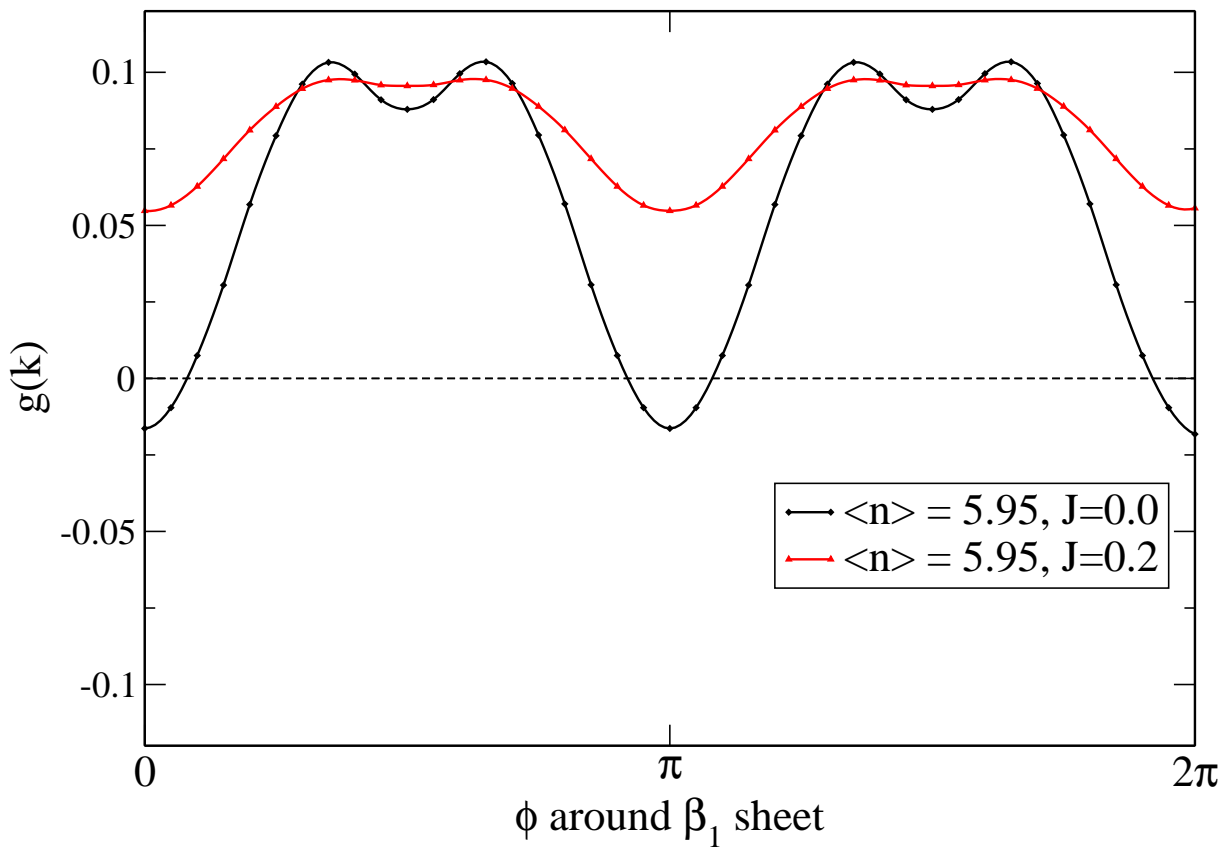}
\caption{The gap function $g(k)$ on the $\beta_1$ pocket for
$n=5.95$, $\bar J=0$ (black diamonds) and
$n=5.95, \bar J=0.2$ (red triangles).
Here the angle $\phi$ is measured from the $k_x$-axis.}
\label{fig:evector_vs_angle_2}
\end{figure}

We return to the role of the Hund's rule exchange $\bar J$ in stabilizing a nodeless
state. The main effect, i.e. the lifting of the nodes as $\bar J$ is
turned on, is illustrated in Fig.~\ref{fig:evector_vs_angle_2}.
There are two reasons this occurs.
First, as discussed in appendix \ref{appendix},
the intra-orbital $\Gamma$'s are controlled by $\bar U$ and $\bar J$
(see Eqs.\ref{eq:chi1111} and \ref{eq:det}).
For a fixed $\bar U$, increasing $\bar J$ drives the system closer to the
instability and thus enhances the leading peak in the RPA susceptibility. However,
Eq.~(\ref{eq:fullGamma}) also contains the term {
\bf ${1\over 2} (\bar U^s + \bar U^c)$}, which
involves only the bare interactions rather than the RPA-enhanced susceptibilities.
This term contributes to the intra-band Coulomb interaction (which favors
anistropic pairing on the electron sheets) and scales, in the dominant orbital
channels, with $\bar U$ rather than $\bar J$. Thus increasing $\bar J$ increases
the importance of the $\chi^{RPA}_1$ term in Eq.~(\ref{eq:fullGamma}) relative to the
intraband Coulomb interaction favoring the isotropic state.

\begin{figure}[ht]
\includegraphics[width=0.9\columnwidth,clip]{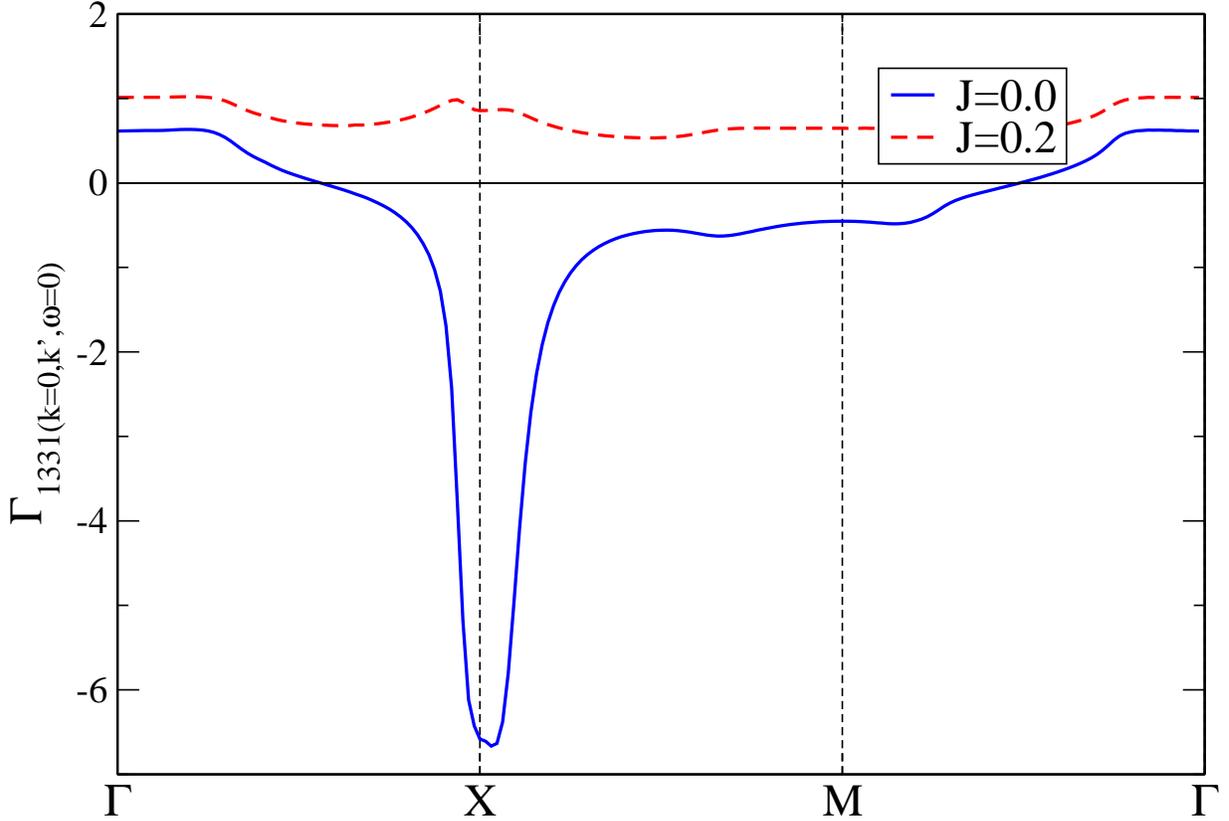}
\caption{The inter-orbital pair scattering vertex $\Gamma_{1331}(\q)$ along
high symmetry directions for $n=5.95$ with parameters $\bar U=1.3$ and  $\bar J=0.0, 0.2$.}
\label{fig:1331}
\end{figure}

Secondly, we find that inter-orbital $d_{xy}$ to $d_{xz}$ pair scattering, represented
by $\Gamma_{1331}$, plays an important role in stabilizing the isotropic state
when a Hund's rule exchange is present.  If $\bar J=0$, Fig. \ref{fig:1331} shows that
the pair scattering has an {\it attractive} peak at $(\pi,0)$ which tends to induce nodes on
the $\beta$ Fermi surface.  If $\bar J>0$, on the other hand, this attractive
peak changes sign, as
discussed in Sec. \ref{sec:4} and in the appendix, 
and a nodeless state is stabilized by the dominant intra-orbital
pairing vertices.

Finally, we note that in Graser et al.\cite{s_graser_08}, upon electron doping
a $d$-wave solution overtakes the anisotropic $s$-wave one. Here we find that
upon strong hole doping ($\sim 8\%$), the spin-fluctuation approximation also
leads to a $d$-wave solution.

\section{Broken spin rotational invariance}
\label{sec:4}

In leading order, the strength of the inter-orbital pair scattering is determined by $U'$ and $J'$.
For SRI parameters $\bar U'=\bar U-\bar J-\bar J'$  the intra-orbital
pair scattering tends to dominate the pairing interaction as we have seen in the
previous section. However, as noted by Zhang et al.\cite{j_zhang_09}, the actual interaction
parameters appropriate for the Fe-superconducting materials need not be SRI.
In this case, inter-orbital pair scattering may play a more important role in
determining the momentum dependence of the gap function $g(\k)$.  In addition,
using non-SRI parameters we can separately explore the roles of the exchange coupling
$\bar J$ and the pair hopping $\bar J'$ interactions on the structure of the pairing
interactions and the gap. 
Here,
for a filling $n=5.95$, we 
hold $\bar U$ and $\bar U'$ fixed and examine the roles of $\bar J$ and
$\bar J'$
on the nodal-gapped transition on the $\beta_1$ pockets near $(\pi,0)$.  The discussion below
applies also to the $\beta_2$ pocket near $(0,\pi)$ with orbital states rotated by 90$^\circ$.

\begin{figure}[ht]
\vspace{5mm}
\includegraphics[width=0.5\columnwidth]{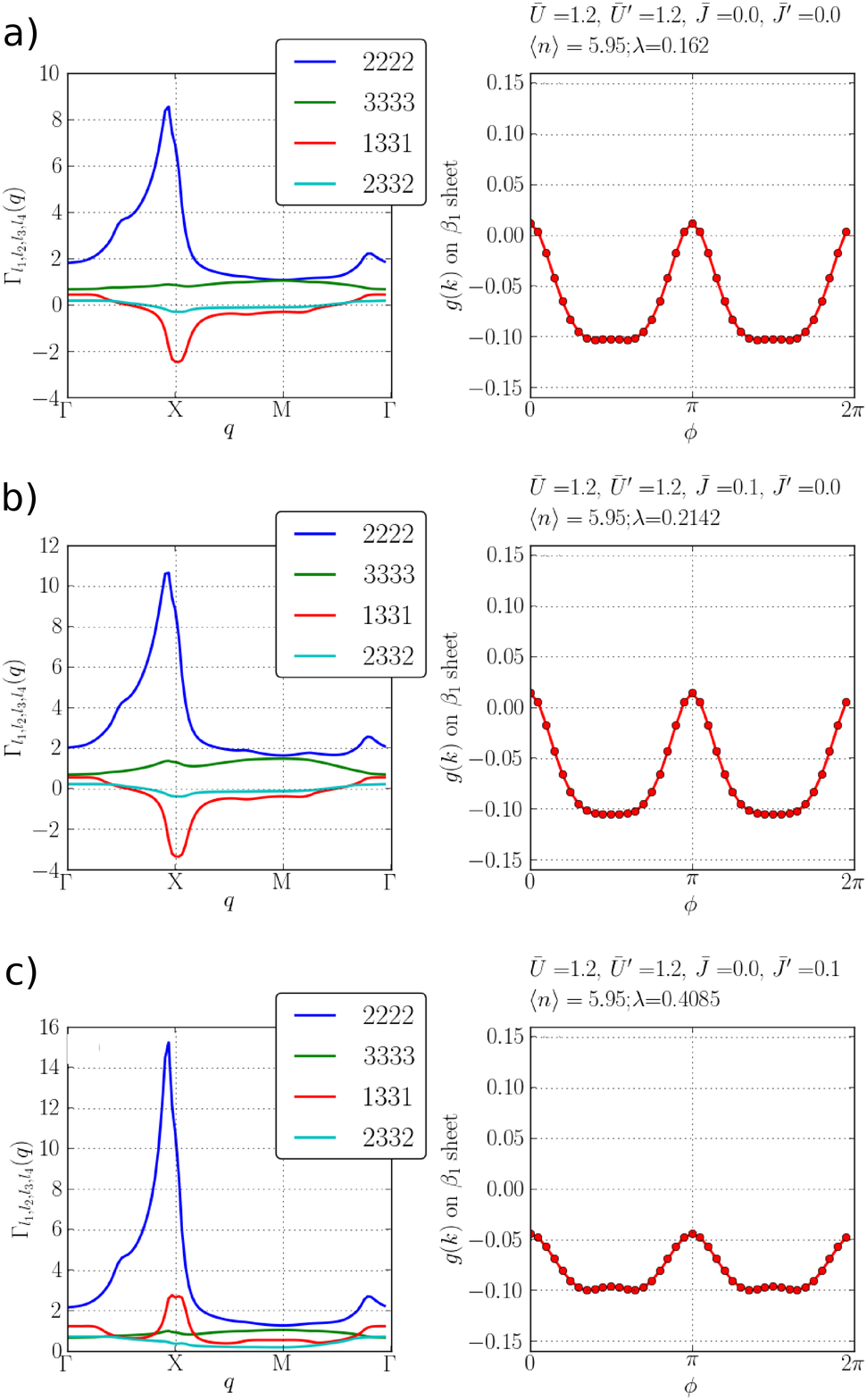}
\caption{Left: orbital pairing vertices, and right:
gap function on the $\beta_1$ sheet, for $\bar U=1.2$, $\bar U'=1.2$ and $n=5.95$.
Cases shown are $\bar J=\bar J'$=0  (a, $\lambda=0.10$); $\bar J=0.1, \bar J'=0$ (b, $\lambda=0.21$); $\bar J=0,\bar J'=0.1$ (c,$\lambda=0.41$).}
\label{fig:nsri_pairing}
\end{figure}

In Fig. \ref{fig:nsri_pairing} we plot both the various 
relevant orbital pairing vertices along high symmetry directions in momentum space, 
as well as the leading gap function $g(\k)$ on the $\beta_1$ sheet.  
Beginning with the nodal case $\bar J = \bar J' = 0$, we can see that increasing
$\bar J$ only weakly affects the momentum space structure of the gap.
However, note that the $g(\k)$ plotted in Fig. \ref{fig:nsri_pairing} is normalized
and should be scaled by the corresponding eigenvalue $\lambda$ to obtain the
true gap amplitude.
Turning on $\bar J$ indeed
increases the pairing eigenvalue $\lambda$ by causing an overall increase in
the pairing vertices. The largest scattering is provided by $\Gamma_{2222}$ near
$(\pi,0)$,
which is driven up by increasing $\bar J$; this can be understood by observing
that the RPA-renormalized susceptibility is enhanced by increasing $\bar J$
(see Eq. \ref{eq:chi1111}). This clearly enhances the strength of the gap on the
parts of the $\beta_1$ Fermi surface characterized by a strong $d_{yz}$ orbital
content. And, as discussed above, the
repulsive nature of the interaction forces the gap on the $d_{yz}$ sections of the
$\beta_1$ sheet to have the opposite sign of the gap on the $d_{yz}$ sections of
the $\alpha_1$ and $\alpha_2$ sheets. This also applies to the $\Gamma_{1111}$
vertex and corresponding $d_{xz}$ sections of the Fermi surface.
The gap on the remaining portions of the Fermi surface, which are of $d_{xy}$ character,
are left to be determined by the largest other pair scattering
with a $d_{xy}$ component, which for these parameters is $\Gamma_{1331}$.
As discussed in appendix A, the Hund's rule coupling $\bar J$ contributes
to an increase in the intra-orbital repulsive scattering which favors
a nodeless gap, while also (see Eq. \ref{eq:1331_J0}) leading to a more
negative $\Gamma_{1331}$ scattering, which favors a nodal gap. The net
result, as seen in Fig. \ref{fig:nsri_pairing}b), is that the nodes remain
for $\bar J=0.1$ and $\bar J'=0$. Alternatively, for the case in which
$\bar J'=0.1$ and $\bar J = 0$, as shown in Fig. \ref{fig:nsri_pairing}c),
the nodes are "lifted." In this case, $\bar J'$ is sufficient to
overcome the negative contribution of $\bar U'^2 \chi^0_{1331}$.
Here, $\Gamma_{1331}$ has changed sign due to the contribution
of Eq. \ref{eq:1331}, which is stronger than the contribution of $\bar U'$
(Eq. \ref{eq:1331_J0})
due to its resonant structure. The change in sign now favors
the same sign across the whole $\beta$ sheet, which causes the nodes to lift.

\section{Orbital character of hole pocket}
\label{sec:5}

\begin{figure}[ht]
\includegraphics[width=0.85\columnwidth]{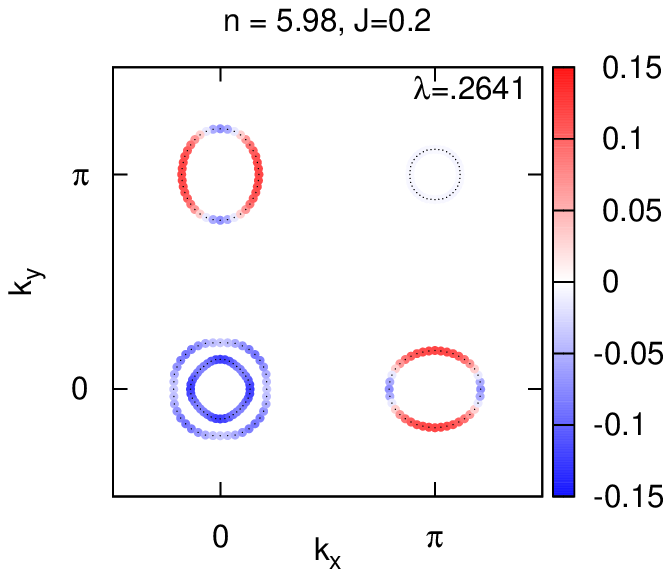}
\caption{The pairing eigenfunction for the hole-doped (x=2\%) compound where the pocket
character has been adjusted to be of $d_{3z^2-r^2}$ type. The interaction parameters
have been chosen as $\bar U=1.3, \bar J=0.2$. }
\label{fig:3z2_pocket}
\end{figure}

We have discussed the appearance of the $\gamma$ hole pocket with doping
in the context of doping by a rigid band shift of the 1111 Fermi surface of
Cao {\it et al.}~\cite{c_cao_08}.  There appear to be various other scenarios in which electronic structure distortions
might occur. For example, as noted by Kuroki {\it et al.}~\cite{k_kuroki_09}, variations in the
As height, which are known to occur in the 1111 family, can tune the size of the $\gamma$
pocket and thereby the pairing itself, independent of doping.  Another effect of the As height
noted by Calder\'on {\it et al.}~\cite{m_calderon_09} is a switching  of two bands near $\Gamma$ such that
the $\gamma$ pocket which occurs upon hole doping is of primarily $d_{3z^2-r^2}$  character rather
than $d_{xy}$.  Within our framework we can imitate  this effect simply by adjusting the size of 
certain tight-binding coefficients associated with the Fe-As hopping in order to create such a $d_{3z^2-r^2}$  pocket
at $\gamma$, and ask what effect this has on the pairing.  As also found by Calder\'on {\it et al.}, the new
$\gamma$ pocket which appears is the only Fermi surface sheet with $d_{3z^2-r^2}$ character, so one may
expect the pairing to be substantially altered relative to the situation with a $\gamma$
pocket of $d_{xy}$ character.  In Fig.~\ref{fig:3z2_pocket}, we see that when the electronic structure is adjusted
to create a $d_{3z^2-r^2}$ pocket instead of a $d_{xy}$,  the pairing eigenfunction reverts to the nodal
$s$-type found in the electron doped case, as expected since the additional $\gamma-\beta$
condensation energy which stabilized the isotropic state has been lost.
In addition, the pairing strength $\lambda$ is substantially reduced.

\section{Effect of surface on pair state}

As noted above, the presence of a hole pocket 
of mainly $d_{xy}$ character around ($\pi$,$\pi$)
in the unfolded zone  causes a nodeless state to be favored over
a nodal one (c.f. Fig.~\ref{fig:eigenvectors}). As pointed out by
Kuroki {\it et al.}~\cite{k_kuroki_08}, this can be accomplished by an
increase of the pnictogen height $h_\mathrm{Pn}$. To provide some
additional insight on the emergence of nodeless behavior, in
particular in ARPES experiments~\cite{l_zhao_08,h_ding_08,t_kondo_08,d_evtushinsky_09,k_nakayama_09,l_wray_08}, we
have performed first-principles calculations on a slab of
\BaFeAs containing 6 FeAs layers.

The calculations were performed using the Quantum-ESPRESSO
package~\cite{pwscf}, which uses a plane wave basis. We used the
Perdew-Burke-Ernzerhof~\cite{jperdew96} exchange-correlation
functionals and ultrasoft pseudopotentials. The use of ultrasoft
pseudopotentials enabled us to utilize an energy cut-off of 40 Ry
for the plane wave basis, while the density cut-off was taken to
be 400 Ry. To determine the positions of the surface ions, two
layers in the middle were kept fixed, and the rest of the atoms
were allowed to relax. We find that the pnictogen height near the
surface and the $c$-axis lattice constant contract, by
about 13\% and 5\%, respectively.

\begin{figure}[ht]
\vspace{5mm}
\includegraphics[width=0.85\columnwidth,angle=0]{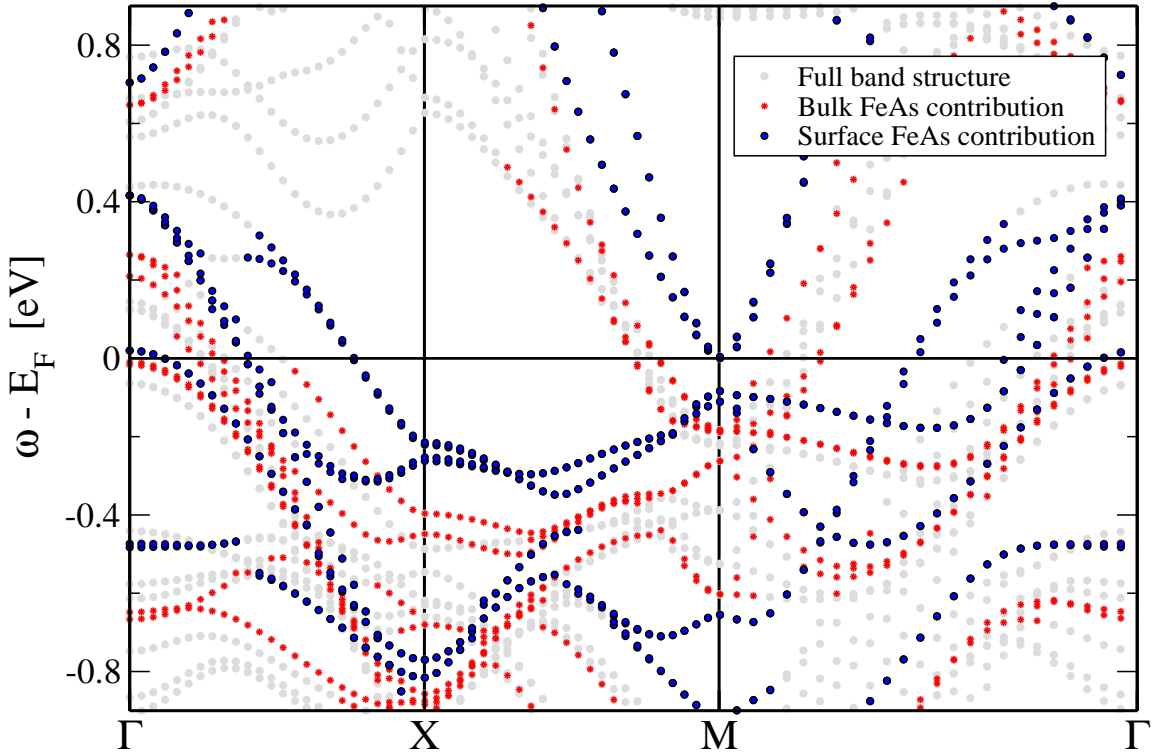}
\caption{The DFT band structure calculated for a
\BaFeAs slab (gray points). The red points show the bulk contributions
from the FeAs layers, while the black points denote the
corresponding surface contributions.}
\label{fig:surface_bands}
\end{figure}

Fig.~\ref{fig:surface_bands} shows the band structure obtained
for the \BaFeAs slab. To determine the origin of the bands, we
projected the band structure on the atomic wavefunctions of the
ions. A particular $\epsilon_k$ is considered to belong to a
certain ion if the projection onto the atom is larger than
50\%. We have verified that the results below do not change
appreciably if the projection threshold is varied.

As can be seen from the figure, the presence of the surface
causes the band just below the Fermi energy in the bulk to rise
slightly, causing the appearance of an additional Fermi surface.
Note that these results are in the folded zone; zone folding
causes the pocket at ($\pi$,$\pi$) to appear at $\Gamma$, and it
is this pocket that is seen due to the surface. For this pocket
to cause a nodeless gap it is necessary that it is of $d_{xy}$
character (or $d_{x^2-y^2}$ in the folded zone), which we have
confirmed (not shown). Thus, it is possible that due to the
sensitive nature of the band structure, the presence
of a surface can cause surface probes to
detect a nodeless gap, even when the bulk gap has nodes.

\section{Conclusions}
\label{sec:6}

One of the striking features of the Fe superconductors is the sensitivity of
the momentum space structure of the superconducting gap to relatively small changes in the
electronic structure.
Indeed, the electronic structure of these materials is quite special. First, they are
semi-metallic with multiple, nearly compensated electron and hole Fermi surfaces.
Secondly, multiple Fe orbitals lie near the Fermi energy. This means that relatively
small changes in the doping or atomic structure can alter the nesting, the orbital
composition of the band states on the Fermi surface, and even the number of $k_z$=0
Fermi surfaces.
The most prominent example of this phenomenon is the appearance of the so-called
$(\pi,\pi)$ hole pocket, studied first by Kuroki {\it et al.}, with hole doping
or pnictogen height.\cite{k_kuroki_09}

Within the framework of an RPA fluctuation exchange approximation, we can understand how
these changes affect the structure of the superconducting gap. For the parameter regime
that we have studied, the leading pairing state has $A_{1g}$ symmetry. It is basically an $s^\pm$
state, but the anisotropy of the gap, and particularly the presence or absence of gap nodes
on the electron Fermi surface sheets is sensitive to the electronic structure.  It is important to note that this type of a 
unconventional nodal state is much more sensitive than for example, the $d_{x^2-y^2}$ pair state of the cuprates, where nodes owe their existence to symmetry; instead, in the Fe-pnictides the nodes appear to be ``accidental", i.e. determined by details of the pair interaction.

As discussed in earlier works, the dominant pairing processes involve intra-orbital scattering. In
this case, the intra-orbital $(\pi,0)$ and $(0,\pi)$ scattering processes lead to a change
of sign between the regions on the $\alpha$ and $\beta$ Fermi surfaces where the $d_{xz}$
and $d_{yz}$ orbital weights are dominant. However, $\beta_1$ to $\beta_2$ intra-orbital
$d_{xy}$ scattering tends to frustrate the isotropic $s^\pm$ state. Furthermore, the effect of the intraband
Coulomb interaction can be reduced in an anisotropic state.
Thus, for light electron doping ($n=6.01$), where there are only the $\alpha$ and $\beta$ Fermi surfaces,
we find that the gap has nodes on the $\beta$ sheets. This anisotropy is further enhanced by the
inter-orbital scattering of $d_{xz}$ ($d_{yz}$) pairs on $\alpha_1$ ($\alpha_2$) to $d_{xy}$ pairs
on the $\beta$ sheets.

In the hole doped case ($n=5.95$), or if the band structure is adjusted, a hole pocket appears
around the \pipi point. For our band structure, this $\gamma$ pocket has $d_{xy}$ character.
Intra-orbital $d_{xy}$ scattering between the $\beta$ and $\gamma$ pockets favors a
more isotropic gap, removing the nodes on the $\beta$ Fermi surfaces. We note that if the band
parameters were such that the orbital character of the $\gamma$ Fermi surface were, for example,
$d_{3z^2-r^2}$ instead of $d_{xy}$, the gap nodes would return to the $\beta$ Fermi surfaces. 
In addition, the pairing strength $\lambda$ is substantially reduced,
reflecting the important role played by the orbital weights
on the Fermi surface.\cite{h_sakakibara_10}
This
illustrates the important role played by the orbital weight. 

We also pointed out that various effects 
in addition to the pnictogen height identified as a key factor by Kuroki {\it et al.} can influence the $\gamma$ pocket.  In particular, the presence of a surface can create such a pocket in a nominally electron-doped system, stabilizing a nodeless state.  This may explain why ARPES experiments have reported quasi-isotropic gaps in these systems.

We have investigated in some detail the way in which the various single-site interaction parameters $\bar U$, $\bar U'$, $\bar J$ and $\bar J'$ influence the different types of
pair scattering processes.  The RPA technique is crude, but it allows analytical insights into questions of this type.  We  have been therefore able to trace the effects of varying these parameters to the strength of the relevant orbital pair scattering vertices at appropriate nesting vectors, which in turn determines not only the overall magnitude of the pairing strength (i.e., $T_c$), but also the anisotropy of the gap on the Fermi surface.  It is to be hoped that the influence of various changes in crystal structure, morphology, etc. on the pair state and transition temperature may now be better understood through this type of analysis.

\acknowledgments{
The authors would like to acknowledge B. A. Bernevig, J. Hu, A. Chubukov, H. Ikeda, I. Mazin, R. Thomale and F. Wang for helpful
comments and discussions. This work is supported by DOE DE-FG02-05ER46236 (PJH, AK) and by DOE/BES
DE-FG02-02ER45995 (HPC, AK). SG acknowledges support by the DFG through SFB 484 and TRR 80,
and DJS and TAM acknowledge the Center for Nanophase Materials Sciences, which is sponsored
at Oak Ridge National Laboratory by the Division of Scientific User Facilities, U.S. Department
of Energy.}

\appendix
\numberwithin{figure}{section}
\section{The scattering vertices}
\label{appendix}

\begin{figure}[ht]
\vspace{5mm}
\includegraphics[width=1\columnwidth]{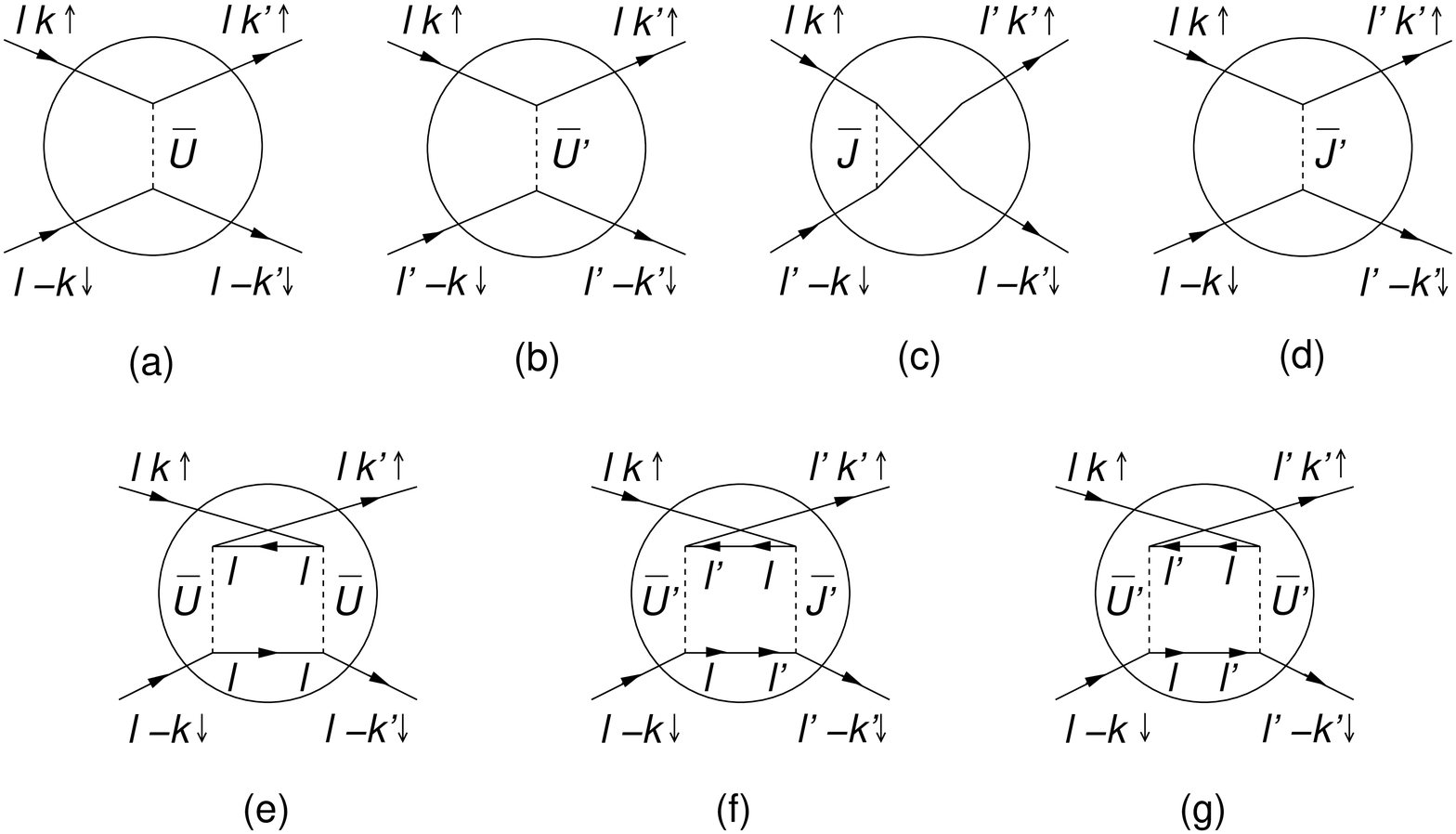}
\caption{The first order (a-d) and some second order (e-g) scattering vertices
corresponding to intra- (a,e), inter- (d,f), and mixed-orbital (b,c,g) scattering processes.  }
\label{fig:ScatteringVertices}
\end{figure}

The basic scattering vertices for the multiorbital Hubbard model
are shown in Fig.~\ref{fig:ScatteringVertices}. Here, as noted in the introduction,
 we use a notation
in which an orbital index $l=(1,2,\dots,5)$ denotes the Fe-$d$ orbitals
($d_{xz}$, $d_{yz}$, $d_{xy}$, $d_{x^2-y^2}$, $d_{3z^2-r^2}$). As seen,
in lowest order, there are intra-orbital (a), inter-orbital (d) and
mixed-orbital (b) and (c) pair scattering processes. Some second order
contributions are shown in Fig.~\ref{fig:ScatteringVertices} (e)-(g).
The orbital matrix elements for $\k$ and $-\k$ states on the Fermi surface
favor pairs which are formed from electrons in the same orbital state.
Thus in spite of the fact that the mixed-orbital scattering can be
significant, its contribution to the pairing interaction is negligible
and the intra- and inter-orbital
scattering processes give rise to the superconducting pairing~\cite{j_zhang_09}.
The relative $l$-orbital contribution to the Bloch state $\k$ on the
$\nu^\textrm{th}$ Fermi surface is given by the square of the orbital
matrix element $|a_\nu^l(\k)|^2$. As shown in
Fig.~\ref{fig:fermisurface}, the $l=1$ ($d_{xz}$) and
$l=2$ ($d_{yz}$) orbitals give the main contributions
on the $\alpha$ Fermi surfaces. Likewise the $l=1$ and $l=3$
($d_{xy}$) orbitals contribute to the $\beta_2$ Fermi surface,
while the $l=2$ and $l=3$ orbitals contribute to the $\beta_1$
Fermi surface.
For our tight-binding fit of the Cao {\it et al.} bandstructure~\cite{c_cao_08},
the $\gamma$ Fermi surface (around $M$=$(\pi,\pi)$)
has predominantly $l=3$ ($d_{xy}$) weight. The orbital weight
distribution favors $(\k,-\k)$ pairs with similar orbitals so that
both, intra-orbital $(d_{xz},d_{xz})_\alpha \leftrightarrow(d_{xz},d_{xz})_{\beta_2}$,
$(d_{yz},d_{yz})_\alpha \leftrightarrow(d_{yz},d_{yz})_{\beta_1}$
as well as inter-orbital $(d_{xz},d_{xz})_\alpha \leftrightarrow(d_{xy},d_{xy})_{\beta_2}$
and  $(d_{yz},d_{yz})_\alpha \leftrightarrow(d_{xy},d_{xy})_{\beta_2}$
pair scattering processes contribute.
In addition, when the $\gamma$ Fermi surface around $(\pi,\pi)$ is present
there are important intra-orbital $(d_{xy},d_{xy})_\gamma \leftrightarrow
(d_{xy},d_{xy})_{\beta_1}$ and $(d_{xy},d_{xy})_\gamma \leftrightarrow (d_{xy},d_{xy})_{\beta_2}$
pair scatterings.
From Fig.~\ref{fig:ScatteringVertices}, one sees that the first order intra-orbital
scattering processes involve $\bar{U}$ and $\bar{J}$ while the
inter-orbital processes depend upon $\bar{U}'$ and $\bar{J}'$.

In the RPA fluctuation exchange approximation, for the parameter range of interest, the
dominant pairing interaction
is given by Eq.~(\ref{eq:fullGamma}) with
\begin{equation}
\chi_1^{\textrm{RPA}}(\q) = \chi^0(\q) \left[ 1 - U^s \chi^{0}(\q) \right]^{-1} 
\end{equation}
and
\begin{equation}
\chi_0^{\textrm{RPA}}(\q) = \chi^0(\q) \left[ 1 + U^c \chi^{0}(\q) \right]^{-1} 
\end{equation}

\begin{figure}[ht]
\includegraphics[width=1\columnwidth]{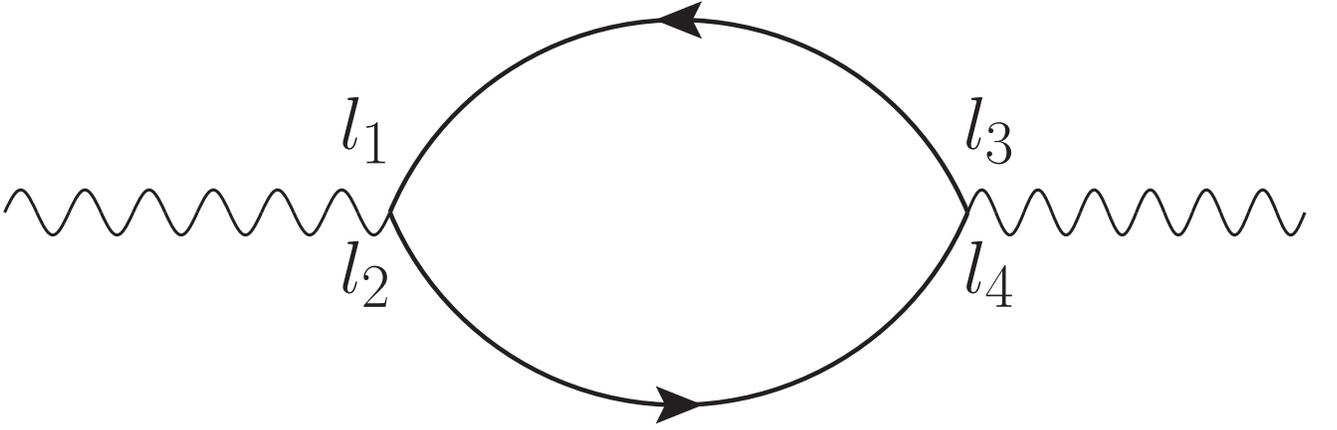}
\caption{Noninteracting susceptibility $\chi^0_{\ell_1,\ell_2,
\ell_3,\ell_4}$ defined in terms of orbital states $\ell_i$ of incoming
and outgoing electrons.}
\label{fig:chi0_diagram}
\end{figure}

For the 5-orbital problem $U^s$, $U^c$ and $\chi^0$ are $25\times 25$ matrices in an
$(\ell_1\ell_2)$ basis with
\begin{eqnarray}
U_{\ell_1 \ell_2 \ell_3 \ell_4}^s &= \left\{
\begin{array}{l l}
\bar{U}\ ,  & \ell_1=\ell_2=\ell_3=\ell_4 \\
\bar{U}^\prime, & \ell_1=\ell_3 \neq \ell_2=\ell_4 \\
\bar{J}\ , & \ell_1=\ell_2  \neq \ell_3=\ell_4 \\
\bar{J}^\prime, & \ell_1=\ell_4 \neq \ell_2=\ell_3
\end{array}
\right. \\
U_{\ell_1 \ell_2 \ell_3 \ell_4}^c &= \left\{
\begin{array}{l l}
\bar{U},  & \ell_1=\ell_2=\ell_3=\ell_4 \\
-\bar{U}^\prime +2\bar{J}, & \ell_1=\ell_3 \neq \ell_2=\ell_4 \\
2\bar{U}^\prime - \bar{J}, & \ell_1=\ell_2  \neq \ell_3=\ell_4 \\
\bar{J}^\prime, & \ell_1=\ell_4 \neq \ell_2=\ell_3
\end{array}
\right.
\end{eqnarray}
and
\begin{eqnarray}
\chi_{\ell_1 \ell_2 \ell_3 \ell_4}^0 (\q) &=& -\frac{1}{N} \sum_{\k,\mu \nu}
\frac{a_\mu^{\ell_4}(\k) a_\mu^{\ell_2 *}(\k) a_\nu^{\ell_1}(\k+\q) a_\nu^{\ell_3 *}(\k+\q)}{
E_\nu (\k+\q) - E_\mu(\k)} \nonumber \\
&& \times \left[ f(E_\nu (\k+\q)) -f(E_\mu (\k)) \right]
\label{eq:spinsusceptibility}
\end{eqnarray}
Here $\mu$ and $\nu$ are summed over the band indices and
$f$ is the usual Fermi function. We have typically taken the temperature
$T=0.02$ eV. The orbital indexing convention for the susceptibility
is illustrated in figure \ref{fig:chi0_diagram}.

\begin{figure}[ht]
\vspace{5mm}
\includegraphics[width=0.95\columnwidth]{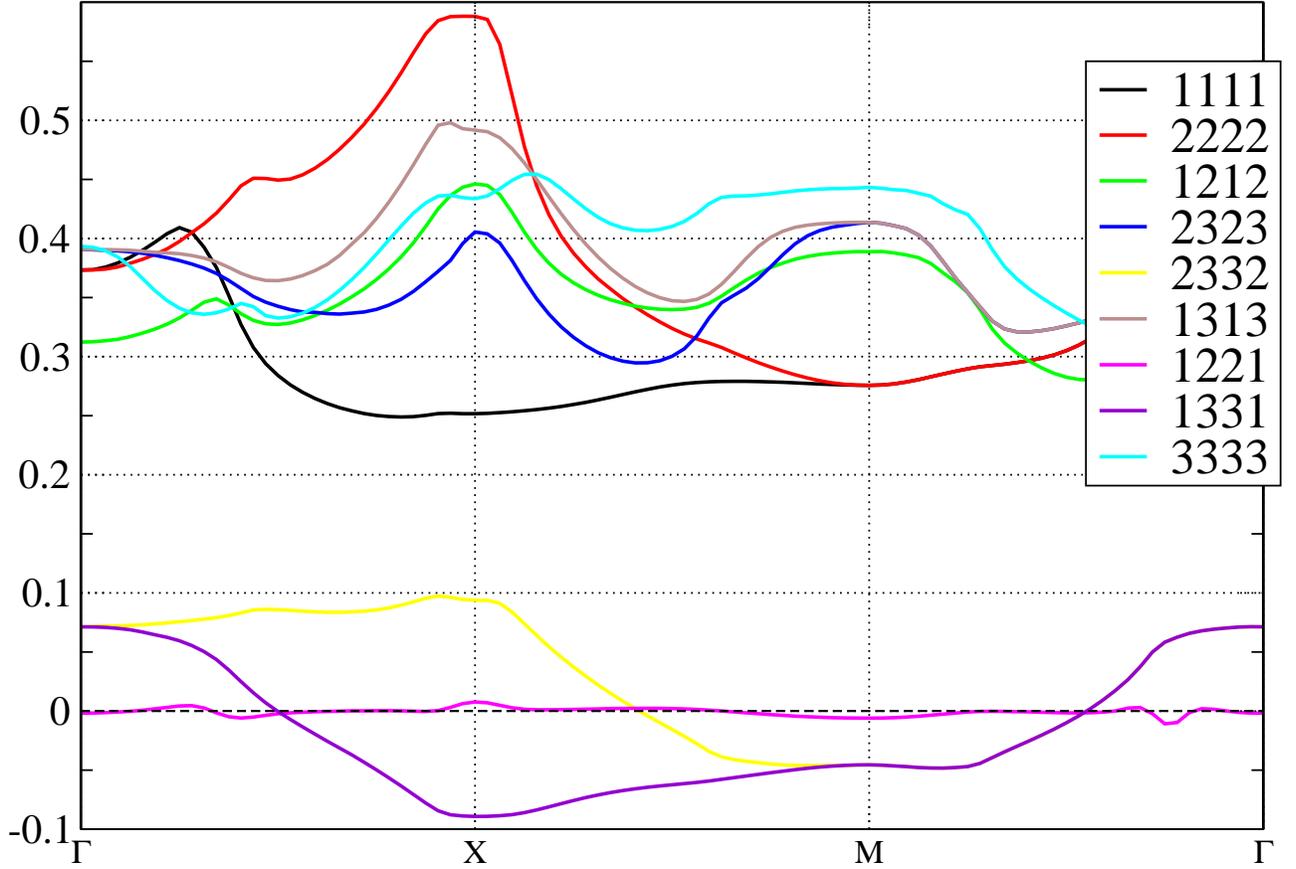}
\caption{The non-interacting susceptibilities $\chi^0_{\ell_1 \ell_2 \ell_3 \ell_4}$ for
$n = 6.01$.}
\label{fig:Chi0}
\end{figure}

As seen in Fig.~\ref{fig:fermisurface}, three orbitals
1 ($d_{xz}$), 2 ($d_{yz}$), and 3 ($d_{xy}$) have significant weight
on the Fermi surfaces. Therefore in this case,
the interaction matrix $U^s$ reduces to the 9$\times$9
matrix shown in Table~\ref{tab:interaction_matrix}.
Furthermore, as seen in Fig.~\ref{fig:Chi0},
the bare off-diagonal elements of the susceptibility
involve single-particle propagators projected on different orbitals
which makes them smaller than the diagonal elements.
\begin{table}
 \begin{center}
 \begin{tabular} {c|ccccccccc}
   & 11 &22 & 33  & 12  & 21  & 13  & 31  & 23  &32   \\\hline\\
    11 & $\bar U$  &  $\bar J$& $ \bar J$ &   &   &   &   &   &   \\
    22 & $\bar J$ &  $\bar U$ &  $\bar J$ &   &   &   &   &   &   \\
    33 &  $\bar J$  &  $\bar J$ & $\bar U$  &   &   &   &   &   &   \\
    12   &   &   &   & $\bar U'$  & $\bar J'$  &   &   &   &   \\
    21   &   &   &   & $\bar J'$  &  $\bar U'$ &   &   &   &   \\
    13  &   &   &   &   &   & $\bar U'$  &$\bar J'$   &   &   \\
    31 &   &   &   &   &   & $\bar J'$  & $\bar U'$  &   &   \\
    23 &   &   &   &   &   &   &   & $\bar U'$  & $\bar J'$  \\
    32   &   &   &   &   &   &   &   &  $\bar J'$ & $\bar U'$  \\
 \end{tabular}
 \end{center}
 \caption{Interaction matrix $U^s$ in the reduced 1 ($d_{xz}$),
  2 ($d_{yz}$), 3 ($d_{xy}$) basis.}
 \label{tab:interaction_matrix}
\end{table}

Keeping only the diagonal terms in $\chi_0$, it is
straightforward to evaluate the RPA susceptibility matrix.
For example,
\begin{eqnarray}
\left(
\begin{array}{cc}\chi_{1313} & \chi_{1331}\\\chi_{3113} & \chi_{3131}
\end{array}\right)^{\rm RPA}
& = & \left(
\begin{array}{cc}\chi_{1313}^0 & 0\\0 & \chi_{3131}^0
\end{array}\right)  \\
& & \times \left(
\begin{array}{cc}1-\bar U'\chi_{1313}^0 & -\bar J'\chi_{1313}^0\\-\bar J'\chi_{3131}^0 & 1-\bar U'\chi_{3131}^0
\end{array}\right)^{-1} \nonumber
\end{eqnarray}
Using the fact that $\chi_{1313}^0 = \chi_{3131}^0$, and setting $\chi_{1313}^0=\chi_{13}^0$, we have
\begin{eqnarray*}
\chi^{\rm RPA}_{1313} &=& {1\over 2} \chi_{13}^0 \left( {1\over 1-(\bar U'+\bar J')\chi_{13}^0}
+ {1\over 1-(\bar U'-\bar J')\chi_{13}^0}\right)  \\
\chi^{\rm RPA}_{1331} &=& {1\over 2} \chi_{13}^0  \left( {1\over 1-(\bar U'+\bar J')\chi_{13}^0}
- {1\over 1-(\bar U'-\bar J')\chi_{13}^0}\right) 
\end{eqnarray*}
In the same way, from the upper left 3$\times$3 part of the interaction matrix one obtains for example
\begin{align}
\chi_{1111}^{RPA}=\frac{\chi_{1}^{0}\left(\left(1-\bar{U}\chi_{2}^{0}\right)\left(1-\bar{U}\chi_{3}^{0}\right)-\bar{J}\chi_{2}^{0}\chi_{3}^{0}\right)}{D}
\label{eq:chi1111}
\end{align}
where $D$ is the determinant of $1-{U}^{s}\chi^{0}$
and takes the form
\begin{align}
D =& \left(1-\bar{U}\chi_{1}^{0}\right)\left(1-\bar{U}\chi_{2}^{0}\right)\left(1-\bar{U}\chi_{3}^{0}\right) \nonumber \\
 & -\bar{J}^{2}\left(1-\bar{U}\chi_{1}^{0}\right)\chi_{2}^{0}\chi_{3}^{0}-\bar{J}^{2}\left(1-\bar{U}\chi_{2}^{0}\right)\chi_{1}^{0}\chi_{3}^{0} \nonumber\\
 & -\bar{J}^{2}\left(1-\bar{U}\chi_{3}^{0}\right)\chi_{1}^{0}\chi_{2}^{0}-2\bar{J}^{3}\chi_{1}^{0}\chi_{2}^{0}\chi_{3}^{0}
\label{eq:det}
\end{align}
with $\chi^0_a\equiv \chi^0_{aaaa}$.  It is clear from these expressions that, for repulsive interactions and within the current approximation,
while the intra-orbital spin susceptibities depend on $\bar U$ and $\bar J$, the 
inter-orbital susceptibilities depend on $\bar U' \pm \bar J'$.

Although we have only kept the diagonal terms in the bare susceptibility,  due to the
structure of the interaction matrices there are off-diagonal terms in the RPA enhanced susceptibility. From
these components, we focus on the intra- and inter-orbital pair scattering processes, as the mixed-orbital
processes are suppressed by the external matrix elements.

Within this diagonal bare susceptibility approximation,
the inter-orbital pair scattering elements of $\Gamma$ are simple
because they involve only a 2$\times $2 interaction matrix.
For example, the pairing strength for the spin-fluctuation
scattering of a $d_{xz}$ pair to a $d_{xy}$ pair is
\begin{equation}
\Gamma_{1331} = \frac{3}{4} \left[ \frac{(\bar U'+\bar J')^2 \chi^0_{13}}{1-(\bar U'+\bar J')\chi_{13}^0}-
\frac{(\bar U'-\bar J')^2 \chi^0_{13}}{1-(\bar U'-\bar J')\chi_{13}^0} \right].
\label{eq:1331}
\end{equation}
There is a corresponding $\Gamma_{2332}$
element which describes the inter-orbital scattering of $d_{yz}$
pairs from the $\alpha$ Fermi surfaces to the $d_{xy}$ regions
of the $\beta_2$ electron Fermi surface.

Since $\chi_{13}^0$ is associated with the nesting of the
$d_{xz}$ and $d_{xy}$ parts of $\alpha_1$ and $\beta_1$
respectively, it peaks for wave vectors near $(\pi,0)$.
For reasonable values of the pair hopping $J'$, there can be a significant
inter-orbital repulsive pair scattering from the $d_{xz}$ regions
of the hole Fermi surfaces $\alpha$ to the $d_{xy}$ regions of the
$\beta$ Fermi surfaces.
As $\bar J^\prime$ goes to zero, the diagonal approximation for this inter-orbital $d_{xz}$ to $d_{xy}$
pair scattering vanishes linearly with $\bar J^\prime$.
If $J'$ becomes small
so that the neglected off-diagonal component of the susceptibility
matrix $\chi_{1331}^0$ are large compared with $\frac{J'}{U'} \chi_{1313}^0$,
then there will be off-diagonal corrections to $\Gamma_{1313}$.
The leading correction is given by
\begin{equation}
\Gamma_{1331} \simeq \bar U'(\bar U' + 2\bar J) \chi^0_{1331}
\label{eq:1331_J0}
\end{equation}
which gives the attractive interaction at $(\pi,0)$ for 
$\bar J=\bar J'=0$, as shown in Fig. \ref{fig:1331}.
In lowest order the $(\bar U^\prime)^2 \chi^0_{1331}$ contribution comes from the second
order processes as seen in Fig.~\ref{fig:ScatteringVertices}.
As seen in Fig.~\ref{fig:Chi0}, $\chi^0_{1331}$ is negative near
the $X$ point where $\q\simeq (\pi,0)$. 

The intra-orbital pair scattering involves a $3\times 3$ matrix and is in general more
complicated. However, in the diagonal susceptibility approximation, the intra-orbital
$\Gamma$ matrix will have the same determinant in the denominator as given by Eq.~(\ref{eq:det}).
From this, one can see how $\bar U$ and $\bar J$ enter in determining the intra-orbital pair
scattering. In particular, the exchange $\bar J$ couples to the various bare susceptibilities
such that, for example, the peak in $\chi^0_{2222}$ near $X$ is reflected in $\Gamma_{1111}$ and
$\Gamma_{3333}$ as well as
$\Gamma_{2222}$ when $\bar J$ is finite.
To summarize, in the diagonal bare susceptibility approximation,
the leading behavior of the intra-orbital pair scattering is controlled
by $\bar U$ and $\bar J$. Increasing these interactions increase the
repulsive scattering at $\q \sim (\pi,0)$ and $(0,\pi)$. The leading
behavior of the inter-orbital scattering is controlled by
$\bar U'$ and $\bar J'$ and in the diagonal bare susceptibility
approximation vanishes as $\bar J'$ goes to zero. When $\bar J'$ is
small, the dominant contribution to the inter-orbital $d_{xz}$ to
$d_{xy}$ and $d_{yz}$ to $d_{xy}$ scattering varies as
$\bar U'(\bar U' + 2J) \chi_{1331}^0$.

\section{The orbital gauge}
\label{appendixB}
In the following section we want to explicitly show that the choice of a gauge for the 
orbital basis does not affect the physical quantities, although one has to carefully
take the orbital gauge into account when comparing orbital dependent  
interaction parameters, as e.g. the pair hopping term $J'$.
First we define a gauge transformation of the orbital creation and annihilation operators as
\[
\tilde{c}_{i \ell \sigma}=c_{i \ell \sigma}e^{i\phi_\ell},
\,\,\,\tilde{c}_{i \ell \sigma}^{\dagger}=c_{i \ell \sigma}^{\dagger}e^{-i\phi_\ell}
\]
where $c_{i \ell \sigma}^{\dagger}$ creates a particle with spin $\sigma$
in orbital $\ell$ at site $i$. Diagonalizing the non-interacting part of the
Fourier-transformed Hamiltonian
the orbital phases can be absorbed into the matrix elements in the form
\[
\tilde{c}_{\ell \sigma}(k)=\sum_{\nu}\tilde{a}_{\nu}^{\ell}(k) \psi_{\nu \sigma}(k)
\]
where $\psi_{\nu \sigma}^\dagger(k)$ creates now a particle with spin $\sigma$
and momentum $k$ in band $\nu$ and the matrix element $\tilde{a}_{\nu}^{\ell}(k) = e^{i \phi_\ell} a_{\nu}^{\ell}(k)$. 
For the bare multiorbital susceptibility as defined in Eq.~\ref{eq:spinsusceptibility} we thus find the following
transformation
\[
 \tilde{\chi}_{\ell_1 \ell_2 \ell_3 \ell_4}^{0}(q,\omega) = 
\chi_{\ell_1 \ell_2 \ell_3 \ell_4}^{0}(q,\omega)e^{i(\phi_{\ell_1}-\phi_{\ell_2}-\phi_{\ell_3}+\phi_{\ell_4})}
\]
while the spin susceptibility as an observable is gauge invariant
\[
 \tilde{\chi}_{S}(q)=\frac{1}{2}\sum_{\ell_1 \ell_2} 
 \tilde{\chi}_{\ell_1 \ell_1 \ell_2 \ell_2}^{0} (q)=\chi_{S}(q).
\]
If we now proceed to the interaction Hamiltonian Eq.~\ref{H} we note that the pair hopping term 
is not gauge invariant
\[
\bar J'\sum_{i, \ell'\neq \ell} \tilde{c}_{i \ell \uparrow}^{\dagger}
\tilde{c}_{i \ell \downarrow}^{\dagger} \tilde{c}_{i \ell' \downarrow} \tilde{c}_{i \ell' \uparrow}
= \bar J' \sum_{i, \ell'\neq \ell} e^{2i(-\phi_{\ell}+\phi_{\ell'})} c_{i \ell \uparrow}^{\dagger}
c_{i \ell \downarrow}^{\dagger} c_{i \ell' \downarrow} c_{i \ell' \uparrow}
\]
and we can define the gauge transformation for $\bar J'$ as
\[
 \tilde{\bar J'} = e^{2i(-\phi_{\ell}+\phi_{\ell'})} \bar J' 
\]
Using this relation allows us to write the gauge transformed interaction matrices as 
\[
\tilde{U}_{\ell_1 \ell_2 \ell_3 \ell_4}^{s/c} = U_{\ell_1 \ell_2 \ell_3 \ell_4}^{s/c} 
  e^{i(\phi_{\ell_1}-\phi_{\ell_2}-\phi_{\ell_3}+\phi_{\ell_4})}.
\]
It is straightforward to show that this relation also yields the correct gauge
transformation for the RPA enhanced multiorbital susceptibility 
\[
 \tilde{\chi}_1^{RPA}(q) = \tilde{\chi}^0(q) + \tilde{\chi}_1^{RPA}(q) \tilde{U}^s \tilde{\chi}^0(q) 
\]
with
\[
(\tilde{\chi}_1^{RPA})_{\ell_1 \ell_2 \ell_3 \ell_4}(q,\omega) = 
(\chi_1^{RPA})_{\ell_1 \ell_2 \ell_3 \ell_4}(q,\omega)e^{i(\phi_{\ell_1}-\phi_{\ell_2}-\phi_{\ell_3}+\phi_{\ell_4})}
\]
under this transformation, the orbital dependent pairing vertex (Eq. \ref{eq:fullGamma}) transforms as 
\[
\tilde{\Gamma}_{\ell_1 \ell_2 \ell_3 \ell_4}(k,k',\omega) = 
\Gamma_{\ell_1 \ell_2 \ell_3 \ell_4}(k,k',\omega)e^{-i(\phi_{\ell_1}-\phi_{\ell_2}-\phi_{\ell_3}+\phi_{\ell_4})}
\]
and we see that the effective pairing vertex entering the linearized gap equation is gauge invariant.

In the above, we have presented the orbital pairing vertices (in
Figs. \ref{fig:orbital_vertices}, \ref{fig:1331} and \ref{fig:nsri_pairing}) and 
orbital susceptibilities (in Fig. \ref{fig:Chi0}). Both of these quantities
depend on the choice of of orbital gauge. We have chosen to present them in the
basis where (1) the $d_{xz}$ and $d_{yz}$ orbitals are aligned along the Fe-Fe directions,
and (2) none of the orbitals have a purely imaginary phase with respect to any other.
In the notation above, $\phi_\ell = 0$, for all $\ell$
In this basis, $a^l_\nu(-\k) = a^{l*}_\nu(\k)$, so that the
sign of the intra- and inter-orbital pairing vertices accurately reflects their
contribution to the gauge invariant pairing vertex $\Gamma_{ij}(\k,\k')$.

\bibliography{master}

\end{document}